\documentclass{SciPost}

\hypersetup{
    colorlinks,
    linkcolor={red!50!black},
    citecolor={blue!50!black},
    urlcolor={blue!80!black}
}

\usepackage[bitstream-charter]{mathdesign}
\DeclareSymbolFont{usualmathcal}{OMS}{cmsy}{m}{n}
\DeclareSymbolFontAlphabet{\mathcal}{usualmathcal}

\fancypagestyle{SPstyle}{
\fancyhf{}
\lhead{\colorbox{scipostblue}{\bf \color{white} ~SciPost Physics }}
\rhead{{\bf \color{scipostdeepblue} ~Submission }}

\fancyfoot[C]{\textbf{\thepage}}
}

\usepackage{braket}
\usepackage{tikz}
\usepackage[normalem]{ulem}

\makeatletter
\usepackage[T1]{fontenc}
\usepackage{amsmath,graphicx}
\usepackage{color}
\usepackage{cancel}
\newcommand{\bfk}{\mathbf{k}}

\newcommand{\hdimer}{\begin{tikzpicture}[baseline=0.0ex,scale=0.25]
  \draw[line width=0.5pt] (0,0) rectangle (1,1);
  \draw[line width=2pt] (0,0) -- (1,0);
  \draw[line width=2pt] (0,1) -- (1,1);
\end{tikzpicture}}

\newcommand{\vdimer}{
\begin{tikzpicture}[baseline=-0.0ex,scale=0.25]
  \draw (0,0) rectangle (1,1);
  \draw[line width=2pt] (0,0) -- (0,1);
  \draw[line width=2pt] (1,0) -- (1,1);
\end{tikzpicture}}

\newcommand{\violI}{
\begin{tikzpicture}[baseline=-0.0ex,scale=0.25]
  \draw (0,0) rectangle (1,1);
  \draw[line width=2pt] (0,0) -- (0,1);
  \draw[line width=2pt] (0,0) -- (1,0);
\end{tikzpicture}}

\newcommand{\violII}{
\begin{tikzpicture}[baseline=-0.0ex,scale=0.25]
  \draw (0,0) rectangle (1,1);
  \draw[line width=2pt] (0,0) -- (0,1);
  \draw[line width=2pt] (1,1) -- (0,1);
\end{tikzpicture}}

\newcommand{\violIII}{
\begin{tikzpicture}[baseline=-0.0ex,scale=0.25]
  \draw (0,0) rectangle (1,1);
  \draw[line width=2pt] (1,1) -- (1,0);
  \draw[line width=2pt] (0,0) -- (1,0);
\end{tikzpicture}}

\newcommand{\violIV}{
\begin{tikzpicture}[baseline=-0.0ex,scale=0.25]
  \draw (0,0) rectangle (1,1);
  \draw[line width=2pt] (1,1) -- (0,1);
  \draw[line width=2pt] (1,1) -- (1,0);
\end{tikzpicture}}

\newcommand{\cstar}{
\begin{tikzpicture}[baseline=0ex,scale=0.25]
    \draw (0,0.5) -- (1,0.5);
    \draw (0,0.5) -- (0.5,1);
    \draw (0,0.5) -- (0.5,0);
    \draw (0.5,0) -- (0.5,1);
    \draw (1,0.5) -- (0.5,1);
    \draw (1,0.5) -- (0.5,0);
\end{tikzpicture}}

\newcommand{\change}[1]
{#1}

\usepackage{color}
\usepackage{soul}

\makeatother

\usepackage{babel}
\begin{document}

\pagestyle{SPstyle}

\begin{center}{\Large \textbf{\color{scipostdeepblue}{
Obstructed Cooper pairs in flat band systems - weakly-coherent superfluids and exact spin liquids\\
}}}\end{center}

\begin{center}\textbf{
Tamaghna Hazra\textsuperscript{1$\star$},
Nishchhal Verma\textsuperscript{2} and
J\"org Schmalian\textsuperscript{1,3}
}\end{center}

\begin{center}
{\bf 1} Institute for Theory of Condensed Matter, Karlsruhe Institute of Technology,
Kaiserstr. 12, 76131 Karlsruhe, Germany
\\
{\bf 2} Department of Physics, Columbia University, New York, NY 10027, USA
\\
{\bf 3} Institut f\"ur Quantenmaterialien und Technologien, Karlsruher Institut f\"ur Technologie, 76131 Karlsruhe, Germany
\\[\baselineskip]
$\star$ \href{mailto:tamaghna.hazra@kit.edu}{\small tamaghna.hazra@kit.edu}
\end{center}

\section*{\color{scipostdeepblue}{Abstract}}
\textbf{\boldmath{%
Superconductivity in a partially filled flat band presents a vexing conceptual hurdle because the absence of a Fermi surface precludes a weak-coupling regime where one can extend insights from the Bardeen-Cooper-Schrieffer picture of a Fermi surface instability. We approach the strongly correlated problem of flat band superconductivity from the strong coupling limit of local attractive interactions on line-graph lattices, whose non-interacting bandstructures host exactly flat bands
originating from geometric frustration of hopping paths. 
In this limit, the pair kinetic energy which sets the superfluid stiffness is expected to scale inversely with the pair binding interaction.
Here we demonstrate a striking counterexample.
We show that when doped charges propagate on the line-graph of a lattice with strong pairing interaction
and broken time-reversal symmetry,
they bind into \textit{obstructed Cooper pairs} whose motion is frustrated by destructive interference. As a result, the leading-order pair kinetic energy
vanishes identically in the strong-coupling expansion, producing a flat bosonic band of compact localised pair states, zero superfluid stiffness at leading order, and an extensively degenerate many-body ground state manifold.
At quarter filling, the frustrated pair dynamics maps onto a quantum dimer model
which in presence of time-reversal breaking, has
a $d$-wave resonating-valence-bond spin liquid ground state, which becomes exact at the analytically solvable Rokhsar-Kivelson point. The pairing Hamiltonian in this limit thus has a topologically ordered ground state with long-range entanglement and deconfined holon excitations. Interestingly, we find exact compact localised eigenstates and extensive degeneracies in the many-body eigenstates of this emergent dimer model. Our results establish a disorder-free mechanism for interaction-driven localisation, in which strong pairing collapses the kinetic energy of Cooper pairs. 
}}

\vspace{\baselineskip}

\noindent\textcolor{white!90!black}{%
\fbox{\parbox{0.975\linewidth}{%
\textcolor{white!40!black}{\begin{tabular}{lr}%
  \begin{minipage}{0.6\textwidth}%
    {\small Copyright attribution to authors. \newline
    This work is a submission to SciPost Physics. \newline
    License information to appear upon publication. \newline
    Publication information to appear upon publication.}
  \end{minipage} & \begin{minipage}{0.4\textwidth}
    {\small Received Date \newline Accepted Date \newline Published Date}%
  \end{minipage}
\end{tabular}}
}}
}


\vspace{10pt}
\noindent\rule{\textwidth}{1pt}
\tableofcontents
\noindent\rule{\textwidth}{1pt}
\vspace{10pt}


\section{Introduction}
The Meissner response that defines a superconductor is quantified by the ability to sustain a smooth deformation of its many-body ground state in response to an external magnetic field - the superfluid stiffness $D_s$. Defined as the free-energy cost of gradients in the phase of the order parameter $\mathcal{F}=D_s\int_r {\rm d}^dr (\nabla \phi_r)^2$
- the static, long-wavelength current-current response in the transverse limit \cite{scalapino1993}, this 
 has the same energy scale as 
the kinetic energy of Cooper pairs in the presence of a magnetic field. In conventional weak-coupling superconductors, the scale of the pair kinetic energy is set by the electronic bandstructure, for a parabolic dispersion $D_s\propto n/m$ where $n$ and $m$ are the density and effective mass of electrons respectively. 
In the presence of a lattice potential,  $n/m$ is replaced by the second derivative of the kinetic Hamiltonian with respect to the crystal momentum, weighted by the density distribution function $n_k$, i.e. no longer fixed by either the particle number or bare mass.
In the strong-coupling limit of such lattice problems, pairs become asymptotically localised and can only move via a second-order process involving a virtual state with dissociated pairs. The conventional expectation is therefore that the stiffness vanishes in the ultimate strong-coupling limit as the inverse of the pair-binding energy~\cite{alexandrov1981,micnas1990}. Here, we present and analyse a model of interacting fermions where the superfluid stiffness is zero in the leading order of the strong-coupling expansion, and instead vanishes much faster as the third power of the inverse pair binding energy. This behavior is caused by the destructive interference of pair-hopping processes~\cite{bergman2008a}, leading to a significant suppression of the superfluid stiffness of a putative superconducting ground state. The exact ground state in a limiting regime of the strong-coupling Hamiltonian is found to be a 
spin-liquid of resonating valence bonds (RVB). This emerges from an explicit mapping of the physics of strongly-coupled local Cooper pairs to a variant of an exactly-solvable quantum dimer model, providing a bridge between the fields of strong-coupling superconductivity and frustrated magnetism.

\begin{figure}
    \centering
    \includegraphics[width=0.4\linewidth]{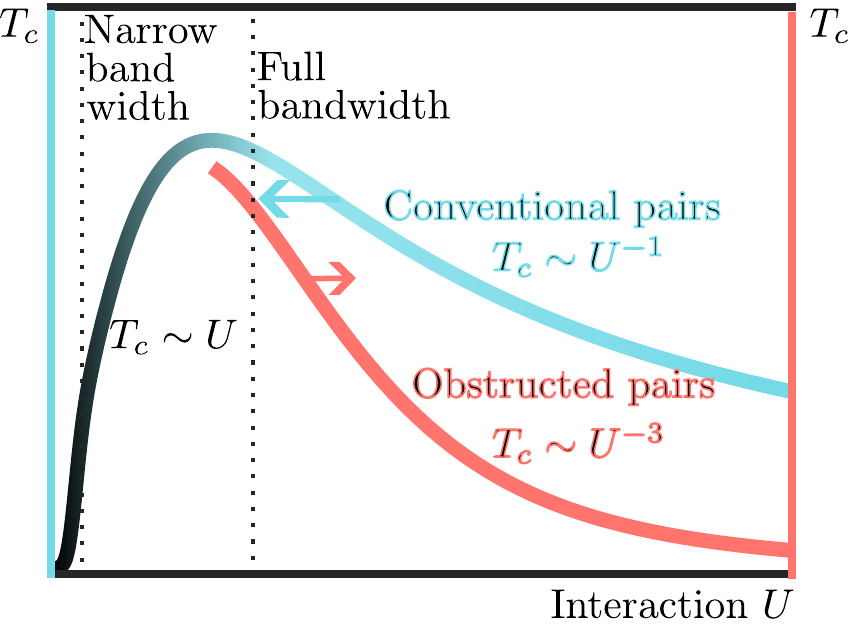}
    \caption{\textbf{Sketch of superconducting $T_c$ and superfluid stiffness in attractive Hubbard models on line-graphs:} When pairing interactions are much stronger than the narrow bandwidth of a flat band, $T_c$ no longer follows the mean-field estimate of the pair-binding energy $T_c\propto \exp(-1/N_0 U)$ where $N_0$ is the large density of states at the Fermi level. Instead, $T_c$ follows the superfluid stiffness which is linear in the Hubbard attraction when its strength is intermediate between the narrow bandwidth and the full bandwidth. When the strength of attraction significantly exceeds the full bandwidth, conventional wisdom predicts a vanishing superfluid stiffness of $\mathcal{O}(U^{-1})$ coming from a second-order pair-hopping process. For some unconventional pairs on line-graphs, dubbed \textit{obstructed pairs}, this leading order contribution is zero and the asymptotic strong-pairing stiffness of a uniform condensate instead scales as $\mathcal{O}(U^{-3})$. This implies that such unconventional superconductivity on a line-graph would have a stiffness and critical temperature orders of magnitude lower than other pairs whose hopping is not obstructed. Note that obstructed pairs are not the leading pairing candidate in general, see Sec 2.}
    \label{fig:FlatBandSC}
\end{figure}

There is much recent interest in the physics of superconductivity in a narrow band that is inherently multi-orbital. 
Experimentally, such flat-band physics has gained relevance across multiple material platforms: Kagome metals and magnets~\cite{wang2023,ghimireTopologyCorrelationsKagome2020,neupertChargeOrderSuperconductivity2022,jiangKagomeSuperconductorsAV3Sb52023}, moir\'e heterostructures of 2D materials~\cite{andreiMarvelsMoireMaterials2021,checkelskyFlatBandsStrange2024}, and in designer quantum simulators of arbitrary graphs using linear microwave resonators~\cite{kollarLineGraphLatticesEuclidean2020}.
Theoretically, it has long been known~\cite{lieb1989,mielke1991,tasaki2020} that exact flat bands are obtained
in line-graph lattices
\footnote{The line graph of a lattice is obtained by taking the links of the lattice as the vertices of the line graph. Any two distinct links that share a common lattice site are considered adjacent and are connected by an edge in the line graph. The adjacency matrix thus constructed has an extensive degeneracy (i.e., a flat band) for any regular lattice that has more links than sites~\cite{mielke1991}. Note that while the vertices of the line graph coincide with the sites of the medial lattice and the terms 'line graph' and 'medial lattice' are sometimes used interchangeably~\cite{roychowdhury2024,chen2025}, the adjacency condition determining the edges of the line graph does not necessarily correspond to nearest-neighbour connectivity of the lattice formed by the edge-centers. Common examples include the checkerboard lattice (the line graph of the square lattice), the kagome lattice (the line graph of the honeycomb lattice), and the pyrochlore lattice (the line graph of the diamond lattice), all of which are known to host exactly flat bands. Similar exactly flat bands are also guaranteed by generalizations of the line-graph construction, as introduced in Ref.~\cite{hazra2026a}. This work focuses on superconductivity in flat-band systems generated by this line-graph paradigm.
We follow the colloquial convention of referring to the checkerboard graph defined by the sites of the medial lattice of the square lattice \textit{and} the adjacency condition of the line graph, as the checkerboard ``lattice''.}
like the Kagome and checkerboard lattice, and in unbalanced bipartite lattices like the Lieb lattice.
There has been significant recent discussion~\cite{peotta2015,julkuGeometricOriginSuperfluidity2016,liangBandGeometryBerry2017,tormaQuantumMetricEffective2018,xie2020,huGeometricConventionalContribution2019,julku2020,iskin2021,hofmann2023a,herzog-arbeitman2022,tovmasyan2016b,herzog-arbeitman2022b} on attractive Hubbard models on such lattices in the interaction regime intermediate between the narrow bandwidth of the flat band and the hopping integrals which set the full bandwidth of the multi-band system (see Fig.~\ref{fig:FlatBandSC}). 
In this regime, the superfluid stiffness has a linear dependence on interaction strength which is upper-bounded~\cite{verma2021,mao2023} and lower-bounded~\cite{peotta2015,xie2020} under some specific conditions by the quantum geometry of the flat band wavefunctions \cite{tormaSuperconductivitySuperfluidityQuantum2022}.
The focus of this work is on the strong-coupling regime of flat band superconductivity in the attractive Hubbard model on a line-graph lattice, specifically for interactions much larger than the tight-binding integrals and the full multi-band bandwidth. 
In this regime, the low-energy effective kinetic Hamiltonian describes local Cooper pairs hopping on the checkerboard lattice.
Complementary to the standard lore in flat band superconductivity that pairs are able to delocalise and \textit{gain} superfluid stiffness from the quantum geometry of flat band wavefunctions, here we find that in the limit of strong coupling,
and in the presence of broken time-reversal,
it is the same non-trivial wavefunctions, now of local Cooper pairs, that lead to perfect localisation and \textit{zero} superfluid stiffness in the leading order of strong-coupling expansions.

Models of hard-core bosons of the type we discuss in this paper have been explored as platforms for exotic quantum spin liquids, potentially useful for topological protection of quantum information\cite{kitaev2003}. In particular, there is significant recent interest in the possibility of using programmable quantum simulators such as Rydberg-atom arrays~\cite{weimer2010,glaetzle2014,samajdar2021,verresen2021} to realise such hard-core bosons as a platform for frustrated magnetism. A remarkable recent advance in this direction~\cite{semeghini2021} was the realisation of $Z_2$ RVB spin liquid on a 219-atom array of Rydberg atoms, each of which simulates a dimer on a link of the Kagome lattice,
with the usual hard-core dimer constraint implemented by tuning the Rydberg blockade radius to the six neighbouring links. 
Here, we show that this physics of quantum dimers and topologically ordered liquids is also realised in the strong-coupling limit of superconductivity in line-graph lattices, whose low-energy effective Hamiltonian in a certain limit maps onto the exactly solvable Rokhsar-Kivelson quantum dimer Hamiltonian. This bridges the physics of strong-coupling superconductivity with frustrated magnetism with the exact result that the ground state of the system in this zero-stiffness limit is a spin-liquid with deconfined holon excitations~\footnote{On non-bipartite lattices like the Kagome lattice and the triangular lattice, this exact ground state is topologically ordered with a $\ln 2$ topological entanglement entropy~\cite{kitaev2006,levin2006,furukawa2007,pei2013,selem2013}}.

Interestingly, our results give credence to the view that even a local, attractive interaction - which would within weak coupling theory always prefer an s-wave pairing state~\cite{brydon2014,scheurer2016}, can give rise to unconventional pairing symmetries in the limit of very strong coupling.
Paradoxically, it is the kinetic energy of local Cooper pairs that favour unconventional pairs that are non-dispersing over delocalised $s$-wave pairs, as we show by diagonalising the emergent pair-hopping effective Hamiltonian in the strong-coupling limit in Section~\ref{sec:boson}. These unconventional pairs fail to delocalise as a result of frustrated hopping pathways on a line-graph~\cite{pudleiner2015}. 
A single pair finds that it gains more kinetic energy by resonating between sublattices, in a compact localised eigenstate of the pair-hopping Hamiltonian, than by delocalising over the lattice in an extended state. 
A compact localised state (CLS)~\cite{maimaiti2017,maimaiti2021,huber2010,morfonios2021} is essentially a molecular orbital that is obstructed from an atomic limit because it has a different sign on different sublattices~\cite{bradlyn2017,cano2018,kruthoffTopologicalClassificationCrystalline2017}. 
We refer to such bosonic CLSs of electron-pairs as obstructed Cooper pairs hereon.

This analysis is done on a lattice relevant for so-called anti-cuprate materials~\cite{kabbour2008,yajima2012a,frandsen2014a,yajima2017a,kelly2025} where the transition metal atoms are located on the links of a square lattice. Specifically, we demonstrate this physics on an attractive Hubbard model on the checkerboard graph, which has an exactly flat band in the limit of nearest-neighbour (NN) and next-nearest-neighbour (NNN) hopping exactly equal. Hopping on the checkerboard graph may be a reasonable starting point for understanding the physics of the anti-cuprate materials if the electrons of the transition metal atoms on the links are the active low-energy degrees of freedom and hopping through the oxygen atoms on the corners dominates over other hopping pathways. While the equality of NN and NNN hopping is a fine-tuned scenario unlikely to be realised exactly in any material platform, the physical insights from our model in this limit can serve as an anchor point for experimental signatures we expect to be observable for a range of parameters which lead to narrow, though not exactly flat, bands. Specifically, we find in this limit (i) an extensive degeneracy of ground states at low-density, (ii) dark localised single-particle wavepackets which are unresponsive to DC electric fields, and in addition in the strong-coupling limit we find (iii) vanishing superfluid stiffness of a putative uniform condensate wavefunction and (iv) a fractionalised phase at quarter-filling that maps to an exactly solvable spin-liquid ground state. 

\begin{figure}
    \centering
    \includegraphics[width=0.8\linewidth]{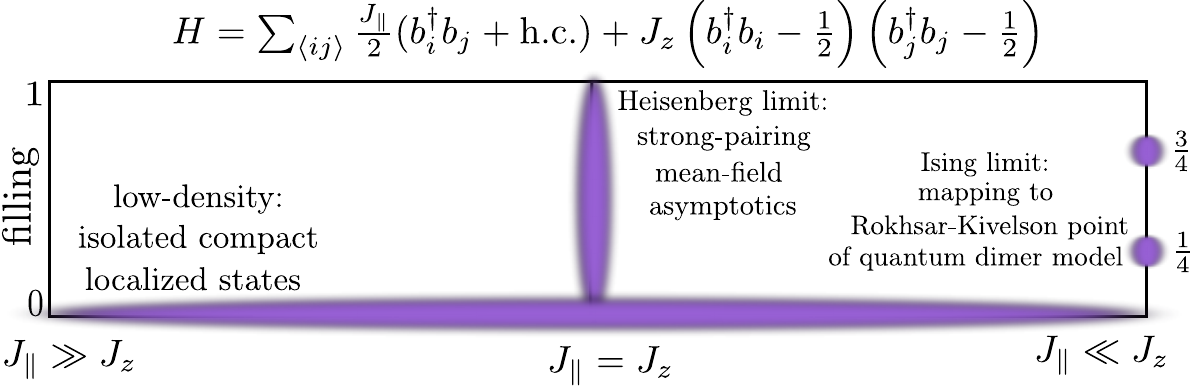}
    \caption{\textbf{Three distinct regimes of a model of hard-core bosons where vanishing superfluid stiffness is established by different techniques.} At low-density, compact localised bosonic states lead to a ground state degeneracy. The Heisenberg limit of equal hopping and repulsion emerges from the strong-coupling expansion of a pairing Hamiltonian which has no leading-order stiffness. Finally, at quarter-filling in the Ising limit, the model maps to a quantum dimer model at the exactly solvable Rokhsar-Kivelson point whose ground state is a RVB spin liquid.}
    \label{fig:phasediagram}
\end{figure}

In the strong-coupling limit of the attractive Hubbard model, the leading-order low-energy effective Hamiltonian describes the physics of hard-core bosons with equally strong nearest-neighbour hopping and repulsion, which maps to a nearest-neighbour Heisenberg interaction of (pseudo)spin-1/2 degrees of freedom on the line-graph. The absence of superfluid stiffness at arbitrary density of bosons then maps to an absence of spin-stiffness against transverse (XY) spin-twists at arbitrary $S_z$ magnetisation for a Heisenberg model on a line-graph. The failure of superfluidity maps to the absence of XY order in spin-exchange Hamiltonians on line-graphs. There are two conceptual anchor-points where the absence of stiffness is intuitively understandable (see Fig.~\ref{fig:phasediagram}). One is the low-density limit of bosons, which corresponds to the nearly $S_z$-polarised limit of spins, where spin-flips are localised in CLS, and fail to delocalise. Another is the limit of Ising anisotropy in the spin-spin interaction where at quarter-filling of bosons, we find an emergent mapping to a quantum dimer model~\cite{kivelsonTopologyResonatingValencebond1987} whose ground state is analytically solvable at the Rokhsar-Kivelson point~\cite{rokhsar1988}. The ground state at this particular filling realises a topologically ordered $d$-wave RVB spin liquid with long-range entanglement, fractionalised spin-0 charge-e holon excitations upon hole-pair doping, and an extensive degeneracy of many-body eigenstates upon electron-pair doping.

\section{Fermionic Model}\label{sec:Hubbard}

Our results hold generally for attractive Hubbard models on line-graphs such as the checkerboard graph, defined by the Hamiltonian $H=H_{\rm kin} + H_U$ where  
\begin{align}
    H_{\rm kin}= \sum_{{\bf k}\sigma}\Psi_{c,{\bf k}\sigma}^{\dagger}h_{{\bf k}\sigma}\Psi_{c,{\bf k}\sigma}
\end{align}
with $\Psi_{c,{\bf k}\sigma}^{\dagger}\equiv\left(\begin{array}{cc}
c_{{\bf k},\frac{\hat{x}}{2}\sigma}^{\dagger} & c_{{\bf k},\frac{\hat{y}}{2}\sigma}^{\dagger}\end{array}\right)$ and 
\begin{equation}
h_{{\bfk}\sigma}\equiv\left(\begin{array}{cc}
(2t_\sigma-\frac{W}{2})\cos k_{x} & 4t_\sigma\cos\frac{k_{x}}{2}\cos\frac{k_{y}}{2}\\
4t_\sigma\cos\frac{k_{x}}{2}\cos\frac{k_{y}}{2} & (2t_\sigma-\frac{W}{2})\cos k_{y}
\end{array}\right)\label{eq:hk}.
\end{equation}
describing fermions hopping on the checkerboard lattice, created by
$c_{{\bf k},\frac{\hat{x}}{2}(\frac{\hat{y}}{2})\sigma}^{\dagger}$ in Bloch states supported on the orbitals on the x(y)-links in each unit cell, and 
\begin{align}
H_{U}=-U\sum_{{\bf r},\hat{\delta}}\left(n_{{\bf r}+\frac{\hat{\delta}}{2}\uparrow}-\frac{1}{2}\right)\left(n_{{\bf r}+\frac{\hat{\delta}}{2}\downarrow}-\frac{1}{2}\right),\label{eq:Uattr}
\end{align}
describing a particle-hole symmetric attractive Hubbard interaction on each link. 
For $t_\sigma>0$, the non-interacting bandstructure has a narrow lower band with width $W$ touching a broad band with width $\sim 8t_\sigma$ at the $M$ point. We will explore two limiting cases of this Hamiltonian. In the $W\to 0$ limit in which the inter-orbital and intra-orbital hopping integrals become exactly equal, the narrow band becomes perfectly flat.
In the strong-coupling limit $U\gg t$, we have all fermions paired up into local Cooper pairs, and no singly-occupied links. 
In the time-reversal symmetric case that we discuss in Sec.~\ref{TR} we have $t_\uparrow=t_\downarrow^*$, the ground state is a conventional $s$-wave superconductor. In Sec.~\ref{TRB} we show that in the opposite limit when $t_{\uparrow}=-t_{\downarrow}^*$ so that the hopping is time-reversal \emph{antisymmetric}, the ground state of the attractive Hubbard model is an unconventional pairing state which has precisely zero superfluid stiffness at the leading order ($\mathcal{O}(1/U)$) in the strong-coupling limit $U\gg t$, where we have all fermions paired up into local Cooper pairs, and no singly-occupied links. For simplicity, we show results for real-valued $t_\sigma$ on the checkerboard lattice, the case of time-reversal breaking hopping that is not $U(1)$-gauge-equivalent to real-valued hopping amounts to fermionic  loop current order and is discussed briefly in Appendix.~\ref{app:TRBhop}. 

\subsection{Time-Reversal Symmetric Case : $t_\uparrow=t_\downarrow$}\label{TR}

\begin{figure}
    \centering
    \includegraphics[width=\linewidth]{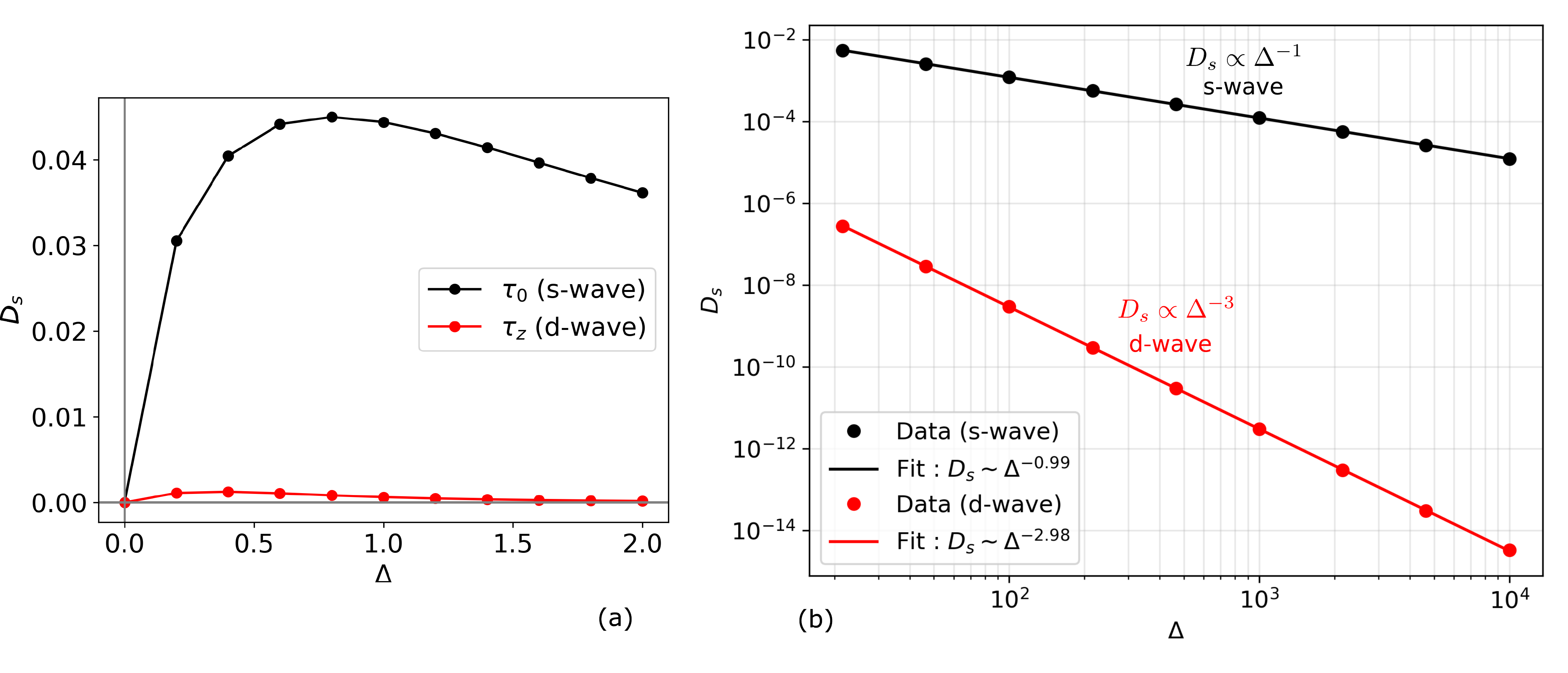}
    \caption{\textbf{Anomalous strong-pairing asymptote of mean-field superfluid stiffness: (Case I) Time-Reversal Symmetric:} Stiffness of a uniform ($q=0$) condensate for sublattice-symmetric (s-wave) and antisymmetric (d-wave) form factors on the checkerboard lattice for fixed density $n=1.95$ for (a) $\Delta<2$ (b) Strong pairing asymptote ($\Delta=2-10000$) of the superfluid stiffness showing the conventional $1/\Delta$ scaling for s-wave pairing, and a faster $1/\Delta^{3}$ scaling for obstructed (d-wave) pairs. The absence of stiffness at the leading order numerically supports the central result of this work. ($\beta=10000,t=1$ on a $40\times40$ lattice)}. 
    \label{fig:meanfield}
\end{figure}

In presence of time-reversal, the attractive Hubbard model has a weak-coupling BCS instability towards $s$-wave pairing, which remains the ground state at arbitrary coupling strength~\cite{hirsch1987,iskin2021}. 
There are two possible choices for a 
lattice-translation-invariant
pairing ansatz, because decoupling the effective Hubbard interaction in the pairing channel yields two on-site pairing order parameters for the two orbitals $\hat\delta\in\{\hat{x},\hat{y}\}$ in each unit cell:
\begin{align}
    -U\sum_{\bf r} n_{{\bf r}+\frac{\hat{\delta}}{2}\uparrow}n_{{\bf r}+\frac{\hat{\delta}}{2}\downarrow} \to 
    \Delta_{\hat{\delta}} \sum_{\bf r} c^\dagger_{{\bf r}+\frac{\hat{\delta}}{2}\uparrow} c^\dagger_{{\bf r}+\frac{\hat{\delta}}{2}\downarrow}.
\end{align}
The symmetric ($\tau_0$) and anti-symmetric ($\tau_z$) combinations of these order parameters have  $s$- and $d_{x^2-y^2}$- wave symmetry respectively.
We calculate the superfluid stiffness of $H=H_{\rm kin} + H_U -\mu (N_{\uparrow}+N_{\downarrow})$ as a function of pairing strength $\Delta/t_{c}$ for fixed electron density $n=1.95$ (See Appendix~\ref{app:superfluid_stiffness} for details) and show (Fig.~\ref{fig:meanfield}) that the obstructed pairs with the sub-leading $\tau_{z}$ (d-wave) form-factor have no stiffness at the leading order $O(1/\Delta)$ in the strong pairing limit - the stiffness vanishes as $D_{s}\propto\Delta^{-3}$. 
In contrast, the uniform pairing condition with $\tau_{0}$ (s-wave) form-factor has the usual strong-coupling asymptote of $D_{s}\propto\Delta^{-1}$ in the strong pairing limit. We have checked that the asymptotic strong-coupling scaling has the same power-law for any density. This motivates us to seek a related Hamiltonian that energetically stabilises the subleading obstructed pairs with $d$-wave form-factor.

\subsection{Time-Reversal Broken Case: $t_\uparrow=-t_\downarrow$}\label{TRB}

\begin{figure}
    \centering
    \includegraphics[width=0.8\linewidth]{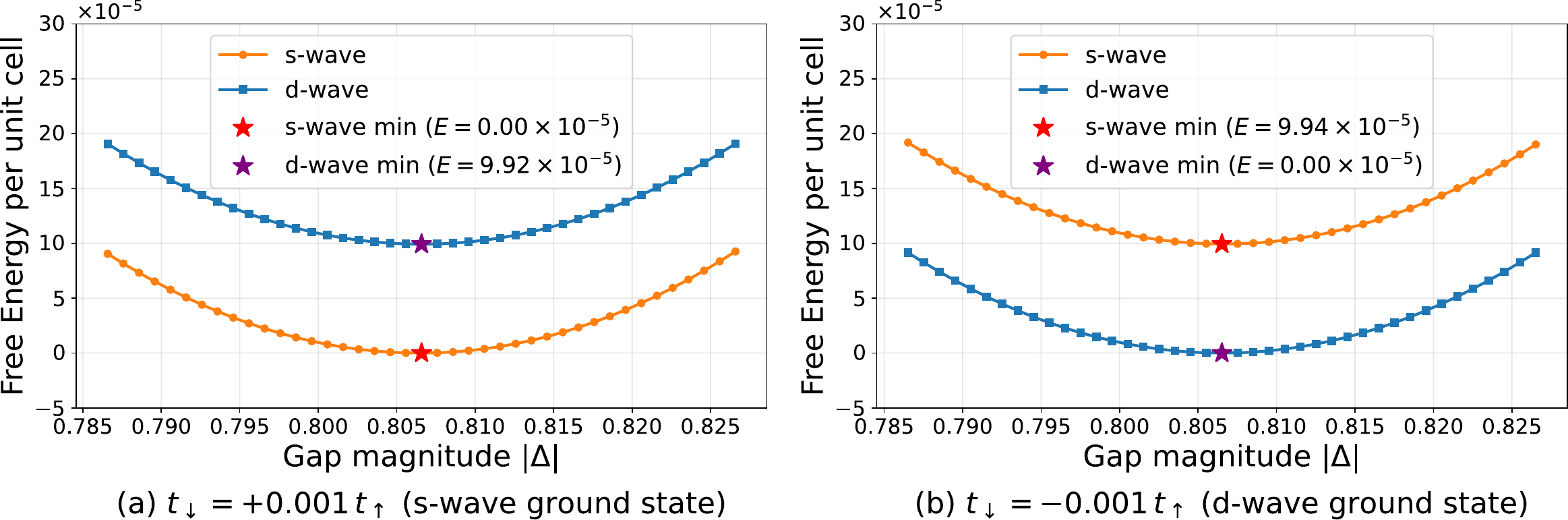}
    \caption{\textbf{Energy Level Crossing:} Landscape of Helmholtz free energy per unit cell zoomed into the minima shows an $s$-wave global minimum at $t_\uparrow t_\downarrow >0$ (a) yielding to a $d$-wave global minimum at $t_\uparrow t_\downarrow <0$ (b). ($\beta=100,t_\uparrow=1,U=4, n=0.2,\mu_\uparrow=\mu_\downarrow$ on a $20\times20$ lattice).} 
    \label{fig:freeEnergyCrossing}
\end{figure}
\begin{figure}
    \centering
    \includegraphics[width=0.5\linewidth]{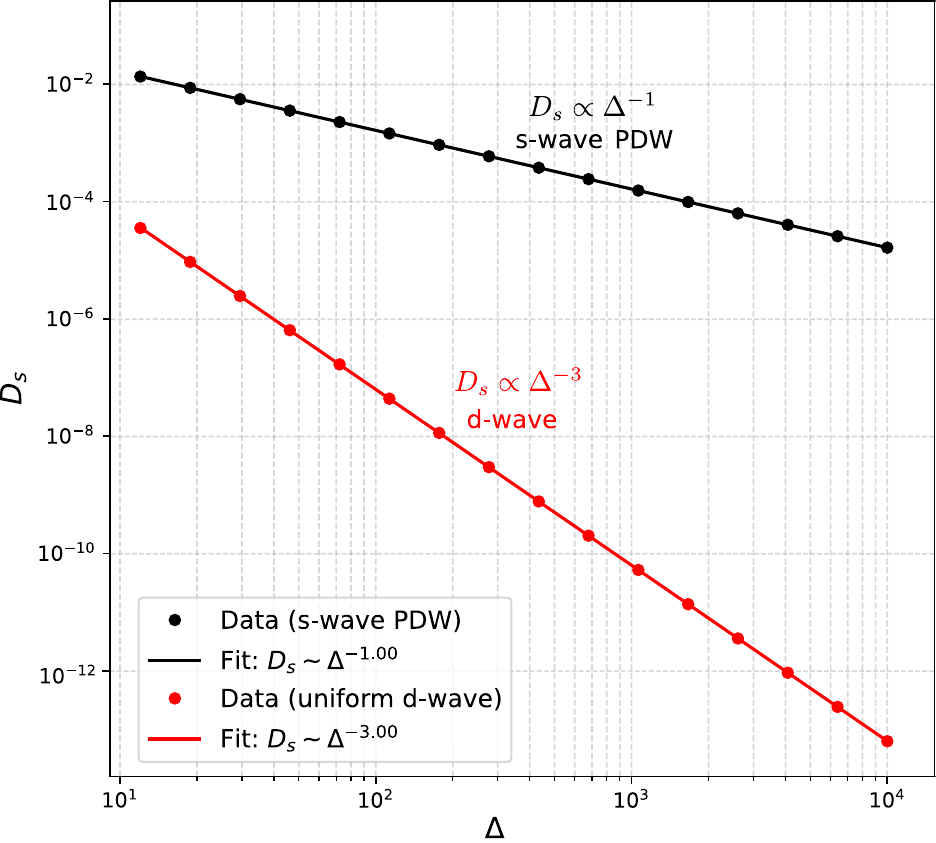}
    \caption{\textbf{Anomalous strong-pairing asymptote of mean-field superfluid stiffness: (Case II:) Time-Reversal Broken:} Stiffness of a uniform ($q=0$) condensate with sublattice-antisymmetric (d-wave, red) form factor and a $q=(\pi,\pi)$ pair-density wave with sublattice-symmetric (s-wave, black) form factor on the checkerboard lattice for fixed densities $n_\uparrow=n_\downarrow=1/10$. The uniform $s$-wave pairing has a negative stiffness (not shown) indicating the instability towards the PDW at $q=(\pi,\pi)$, which shows the conventional $1/\Delta$ scaling, while the uniform d-wave (obstructed) pair condensate shows a faster $1/\Delta^{3}$ scaling. The absence of stiffness at the leading order numerically supports the central result of this work. ($\beta=100,t_\uparrow=1,t_\downarrow=-1$ on a $40\times40$ lattice).} 
    \label{fig:meanfield2}
\end{figure}

Interpolating from $t_\downarrow/t_\uparrow = 1$ to $t_\downarrow/t_\uparrow = -1$, we find a level crossing between the $s$-wave ground state and the subleading $d$-wave gap at $t_\downarrow=0$ beyond which the unconventional $d$-wave superconductor is the stable ground state (Fig.~\ref{fig:freeEnergyCrossing}). Computing the superfluid stiffness (See Appendix~\ref{app:superfluid_stiffness} for details) of $H=H_{\rm kin} + H_U -\mu_\uparrow N_{\uparrow}-\mu_\downarrow N_{\downarrow}$, we find that the stiffness of the now subleading $s$-wave $q=0$ condensate is negative, indicating its instability towards a state with finite center-of-mass momentum pairing. We find the global minimum in the $s$-wave sector at strong coupling to be a pair-density wave (PDW) at ${\bf q}=(\pi,\pi)$.~\footnote{To establish this, we decoupled the interaction $-U\sum_{\bf r} n_{{\bf r}+\frac{\hat{\delta}}{2}\uparrow}n_{{\bf r}+\frac{\hat{\delta}}{2}\downarrow} \to \sum_{\bf r} \Delta_{{\bf r} \hat{\delta}}c^\dagger_{{\bf r}+\frac{\hat{\delta}}{2}\uparrow} c^\dagger_{{\bf r}+\frac{\hat{\delta}}{2}}$ \emph{without} assuming a translation-invariant gap function and solved the full Bogoliubov-deGennes gap equation iteratively with $\Delta_{{\bf q} \hat{x}}=\Delta_{{\bf q} \hat{y}}\neq0$ at all $q$  for  $U=10t_\uparrow,t_\downarrow=-t_\uparrow$.} The relative Helmholtz free energies of the extremal solutions are shown in Table~\ref{tab:free_energy} with the corresponding superfluid stiffness. In Fig.~\ref{fig:meanfield2}, we show that the stiffness of the $q=(\pi,\pi)$ PDW is positive and follows the conventional strong coupling asymptote of $D_s \propto \Delta^{-1}$, while the $q=(0,0)$ $d$-wave condensate shows $D_s \propto \Delta^{-3}$, as in the time-reversal invariant case, also for any density. In contrast to that case, obstructed pairing with no leading-order stiffness is now the ground state.

\begin{table}[h]
    \centering
    \begin{tabular}{|l|c|c|c|c|}
        \hline
        Gap & $\Delta_{\hat{x}/2}/\Delta_{\hat{y}/2}$ & $\mathbf{q}$& Superfluid stiffness $D_s$ & Free energy \\
        \hline
        Uniform d-wave & $-1$ & $(0,0)$ & $6.12 \times 10^{-6}$ & $0$ \\
        s-wave PDW & $+1$ & $(\pi,\pi)$ & $7.59 \times 10^{-3}$ & $3.07 \times 10^{-6}$ \\
        d-wave PDW & $-1$ & $(\pi,\pi)$ & $7.59 \times 10^{-3}$ & $3.07 \times 10^{-6}$~\footnotemark \\
        Uniform s-wave & $+1$ & $(0,0)$ & $-1.52 \times 10^{-2}$ & $7.60 \times 10^{-3}$ \\
        \hline
    \end{tabular}
    \caption{Free energy and superfluid stiffness per unit cell for different pairing ans\"atze on a checkerboard lattice attractive Hubbard model, for $U=100, t_\uparrow=1,t_\downarrow=-1, \beta=10000$ on a $40\times40$ system. Unlike the time-reversal symmetric case of $t_\uparrow=t_\downarrow$, the uniform s-wave condensate has negative superfluid stiffness indicating an instability towards a pair-density wave at $\mathbf{Q}=(\pi,\pi)$. The two PDW ans\"atze are unitarily equivalent\protect\footnotemark[\value{footnote}] and thus have the same free energy.} 
    \label{tab:free_energy}
\end{table}
\footnotetext{In the basis $\Psi_{\mathbf{k}} = (c_{x, \mathbf{k}, \uparrow}, c_{y, \mathbf{k}, \uparrow}, c^\dagger_{x, -\mathbf{k}+\mathbf{q}, \downarrow}, c^\dagger_{y, -\mathbf{k}+\mathbf{q}, \downarrow})^T$, the BdG Hamiltonian matrices for the $s$-wave and $d$-wave PDW states at $\mathbf{q} = (\pi, \pi)$ are exactly related by the unitary transformation $H_d(k_x, k_y) = V H_s(-k_x, k_y) V^\dagger$ with $V = \text{diag}(1, 1, 1, -1)$, corresponding to a $\pi$-phase gauge shift on the down-spin electrons on the $y$-links followed by a spatial reflection $k_x \to -k_x$ that absorbs the odd-parity sign inversion in the inter-orbital hole kinetic terms. Because the Hamiltonians are unitarily equivalent, the free energy is exactly equal up to numerical error.}

We have considered the superfluid stiffness of four variational BCS ground states in the strong-pairing limit. In principle, the strong-coupling limit should conserve separately the number of spin up and spin down electrons with separate Lagrange multipliers $\mu_\uparrow$ and $\mu_\downarrow$, but in practice the energy cost of having unpaired majority-spin electrons drives the system towards $n_\uparrow=n_\downarrow$ at strong-coupling despite the presence of a spin-polarisation in the non-interacting system. For time-reversal invariant attractive Hubbard interaction on a single dispersive band, the existence of a full gap in the fermionic spectrum enables a smooth crossover from the weak-coupling BCS regime with a pairing instability of the Fermi surface to the strong-coupling BEC regime where the Cooper pair condensate acquires a finite superfluid stiffness at temperatures well below their pair-binding energy~\cite{randeria2014}. In the absence of time-reversal symmetry, there is no weak-coupling Cooper instability towards singlet pairing because the pairing interaction must overcome the kinetic energy difference between the time-reversal partners. In an unconventional superconductor with gap nodes, there is no adiabatic crossover from weak to strong coupling and no controlled approximation scheme to extend the results of weak-coupling mean-field theory to the strong-pairing regime. 

Thus, it remains to be shown that the ordering tendencies indicated by mean-field analysis and the absence of superfluid stiffness are borne out in a calculation that is exact in the strong-coupling limit. 
In the next section, we analyse the attractive Hubbard model on the checkerboard graph using a perturbation expansion around the strong-coupling limit. 
We find that when $t_\uparrow t_\downarrow <0$, the kinetic energy of tightly-bound on-site pairs favours the orbitally-antisymmetric $d$-wave form-factor over the orbitally-symmetric $s$-wave form-factor near $Q=0$.
Thus even for attractive Hubbard interactions, which are conventionally expected to have $s$-wave pairing order parameters, the kinetic energy of on-site pairs can stabilise unconventional pairing in the strong-coupling limit when time-reversal symmetry is broken.
This strong-coupling analysis also explains why a leading order contribution to the superfluid stiffness is absent for a condensate of $d$-wave pairs on the checkerboard graph, a result anticipated already from our mean-field analysis (Fig.~\ref{fig:meanfield},~\ref{fig:meanfield2}). 
In the low-density limit, this agrees with ongoing discussions \cite{huhtinen2026interplaylocalglobalquantum,huhtinen2026stabilityflatbandboseeinsteincondensation} on the lack of stability of flat band Bose-Einstein condensates from the perspective of quantum geometry.

\section{Insights from hard-core bosons in the low-density limit}\label{sec:boson}

We derive the low-energy effective Hamiltonian in the strong-pairing limit $U\gg t_\sigma$. The low-energy sector excludes single-occupancy of the link-orbitals, and by a standard Schrieffer-Wolff transformation (See Appendix~\ref{app:SW1} for details), we obtain the effective boson Hamiltonian 
\begin{align}
 H&=H_{{\rm hop}}+H_{NN}
 \label{eq:boson}\\
 H_{{\rm hop}}&=\sum_{{\bf k}}\Psi_{b,{\bf k}}^{\dagger}h^b_{{\bf k}}\Psi_{b,{\bf k}}\notag\\
 H_{NN}&=2\sum_{{\bf k}}N_{{\bf k}}^T V^b_{{\bf k}}N_{{\bf k}}\notag
\end{align}
Here $\Psi_{b,{\bf k}}^{\dagger}\equiv\left(\begin{array}{cc}
b_{{\bf k},\frac{\hat{x}}{2}}^{\dagger} & b_{{\bf k},\frac{\hat{y}}{2}}^{\dagger}\end{array}\right)$,
$b_{{\bf r},\frac{\hat{x}}{2}}^{\dagger}\equiv c_{{\bf r},\frac{\hat{x}}{2}\uparrow}^{\dagger} c_{{\bf r},\frac{\hat{x}}{2}\downarrow}^{\dagger}|0\rangle$ describes hard-core bosons with charge -2e,
$N_{{\bf k}}^T \equiv\left(
    \begin{array}{cc} 
        n_{{\bf k},\frac{\hat{x}}{2}}^{b}-\frac{1}{2} & n_{{\bf k},\frac{\hat{y}}{2}}^{b} -\frac{1}{2} 
    \end{array}
    \right)$, 
$n_{{\bf k},\frac{\hat{\delta}}{2}}^{b}=b^\dagger_{{\bf k},\frac{\hat{\delta}}{2}} b_{{\bf k},\frac{\hat{\delta}}{2}}$ describes their particle-hole symmetric density operators.

In \eqref{eq:boson}, $H_{\rm hop}$ describes hard core bosons hopping between links that share a corner site, with
\[h^b_{\bfk}\equiv\left(\begin{array}{cc}
2t_{b}^{0}\cos k_{x}-2\mu & 4t_{b}^{\times}\cos\frac{k_{x}}{2}\cos\frac{k_{y}}{2}\\
4t_{b}^{\times}\cos\frac{k_{x}}{2}\cos\frac{k_{y}}{2} & 2t_{b}^{0}\cos k_{y}-2\mu.
\end{array}\right)\]
with $t_{b}^{\times}=-2t_\uparrow t_\downarrow/U\equiv t_b$ and $t_{b}^{0}=-2(t_\uparrow-W/4)(t_\downarrow-W/4)/U\equiv t_b-W_b/4$, describing inter- and intra-sublattice pair hopping, respectively.
Note that the positive sign of the pair-hopping integral $t_{b}=-2t_\uparrow t_\downarrow/U>0$ is a direct consequence of the broken time-reversal symmetry ($t_\uparrow t_\downarrow < 0$). In a conventional time-reversal invariant attractive Hubbard model, $t_\downarrow = t_\uparrow$, so the pair-hopping integral $t_b$ is negative definite, and conventional $s$-wave pairing is dominant at strong coupling. 

In addition, \eqref{eq:boson} also contains a density-density repulsion between bosons on links that share a corner site, with
\[V^b_{\bfk}\equiv\left(\begin{array}{cc}
2V_{b}^{0}\cos k_{x}-2\mu & 4V_{b}^{\times}\cos\frac{k_{x}}{2}\cos\frac{k_{y}}{2}\\
4V_{b}^{\times}\cos\frac{k_{x}}{2}\cos\frac{k_{y}}{2} & 2V_{b}^{0}\cos k_{y}-2\mu.
\end{array}\right)\]
with $V_{b}^{\times}=(t_\uparrow t_\uparrow + t_\downarrow t_\downarrow)/U\equiv V_b$ and $V_{b}^{0}=((t_\uparrow-W/4) (t_\uparrow -W/4)+ (t_\downarrow-W/4) (t_\downarrow-W/4))/U\equiv V_b-\delta V_b$, describing inter- and intra-sublattice pair repulsion, respectively. In the time-reversal symmetric case where $t_\downarrow = t_\uparrow$ this describes equal strength pair-hopping and pair-repulsion with $V_b = -t_b, \delta V_b=-W_b$. In the time-reversal anti-symmetric case where $t_\downarrow = -t_\uparrow$, we obtain an emergent Heisenberg model as indicated in Fig.~\ref{fig:phasediagram}.

\subsection{Kinetic stabilisation of $d$-wave pairing}

\begin{figure}
\centering
\includegraphics[width=0.8\linewidth]{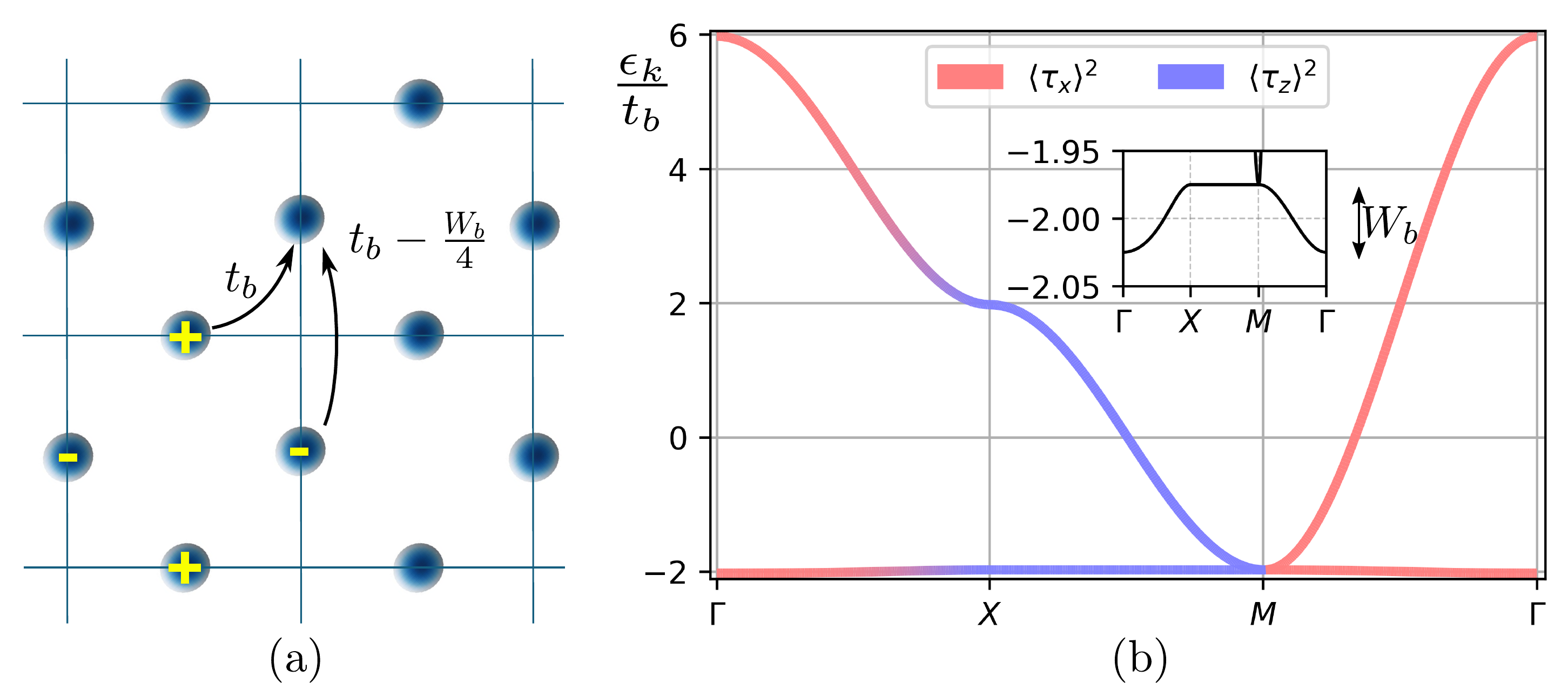} 
\caption{Pair-hopping integrals (a), band structure (b) of the pair-Hamiltonian $H_{\rm hop}$ in Eq.~\eqref{eq:boson} with $W_b=0.05t_{b}$.  Also shown in (a) is a compact localised state (CLS) on the lower left plaquette that is prevented from hopping to the neighbouring plaquette by destructive interference in the limit $W_b\to 0$. The colour in (b) indicates the orbital content of the band eigenfunctions, which winds twice around the band-touching point at $M$.
\label{fig:checkerboard}}
\end{figure}

First, we consider the Hilbert space of a single boson. Diagonalising $H_{\rm hop}$ immediately yields a nearly flat bosonic band with bandwidth $W_b=((t_\uparrow + t_\downarrow)W-W^2)/(2U)$ and minimum at $\Gamma$ touching a dispersive band with bandwidth $\approx 8t_b=-16t_\uparrow t_\downarrow/U$ at the $M$ point (see Fig.\ref{fig:checkerboard}(b))~\footnote{This band-touching is protected by crystalline and time-reversal symmetries, and a time-reversal breaking pair-hopping perturbation $H=4 t_M \tau_y\sin k_x/2 \sin k_y/2$ stabilises a topologically non-trivial condensate at $M$ with a finite Chern number as in \cite{jalali-mola2023}. }. At the $\Gamma$ point, for $t_\uparrow=-t_\downarrow$, the lower (flat) band eigenstate $(1,-1)/\sqrt{2}$ transforms under the $B_{1g}$ irrep of the little group and the upper (dispersive) band eigenstate $(1,1)/\sqrt{2}$ transforms trivially under the $A_{1g}$ irrep~\footnote{For the time-reversal invariant ($t_\uparrow=t_\downarrow$) case, the ordering of flat and dispersive bands is reversed.}.

The kinetic energy difference of $8t_b$ between the two $q=0$ eigenstates has implications for the energy of a uniform condensate of pairs. A uniform $s$-wave condensate costs $8t_b$ more kinetic energy per boson than the condensate of the lowest energy $q=0$ bosons, which is a $d$-wave condensate. 

\subsection{Dark localised eigenstates}
When $W=0$, any superposition of the degenerate Bloch eigenstates $|\Psi_{{\bf  k}{\rm flat}}\rangle$ of $H_{\rm hop}$ is also an exact eigenstate. 
It is possible to construct localised wavepackets $|\Psi_{{\bf R},{\Phi}}\rangle\equiv\sum_{{\bf k}}e^{i{\bf k}\cdot{\bf R}}\Phi_{{\bf k}}|\psi_{{\bf k}<}\rangle$ centered at any site ${\bf R}$ with an arbitrary envelope function $\Phi$. 
Being superpositions of degenerate eigenstates, these remain localised under unitary time-evolution $e^{-iHt}|\Psi_{{\bf R},{\Phi}}\rangle=e^{2it_{b}t}|\Psi_{{\bf R},{\Phi}}\rangle$.
Moreover, in each of the Bloch eigenstates $|\Psi_{{\bf  k}{\rm flat}}\rangle,{\bf k}\neq M$, we show (Appendix~\ref{app:darkloc}) that the linear response to ${\bf {A}}$, a uniform $U(1)$ vector potential, is exactly zero - there is perfect cancellation between the diamagnetic and paramagnetic current response. 
These localised wavepackets have no Drude response to a DC electric field. 

The flat band does not have exponentially localised Wannier functions, due to a non-analytic winding of the Bloch eigenvector at the $M$ point where the flat band touches the dispersive band (see Fig~\ref{fig:checkerboard}(b)). Nevertheless, the $N+1$-fold degenerate manifold of single-boson ground states can be spanned by two extended states which wind around the periodic sample~\cite{bergman2008a}, and $N-1$ compact localised states whose wavefunction takes the form $$|\Psi_{\bf r}^{\rm CLS} \rangle  \equiv \frac{1}{2}\left(b_{{\bf r},\frac{\hat{x}}{2}}^{\dagger}-b_{{\bf r},\frac{\hat{y}}{2}}^{\dagger}+b_{{\bf r}+\hat{y},\frac{\hat{x}}{2}}^{\dagger}-b_{{\bf r}+\hat{x},\frac{\hat{y}}{2}}^{\dagger}\right)|0\rangle,$$ describing localised d-wave pairs that resonate between these four local configurations around a plaquette. Although the amplitude of the pair hopping is finite, these pairs are prevented from hopping onto neighbouring plaquettes by the destructive interference of the phase. Hence they have zero contribution to the long-wavelength current response.  

\subsection{Many-body ground state degeneracy}\label{subsec:MB}
\begin{figure}
    \centering
    \includegraphics[width=0.6\linewidth]{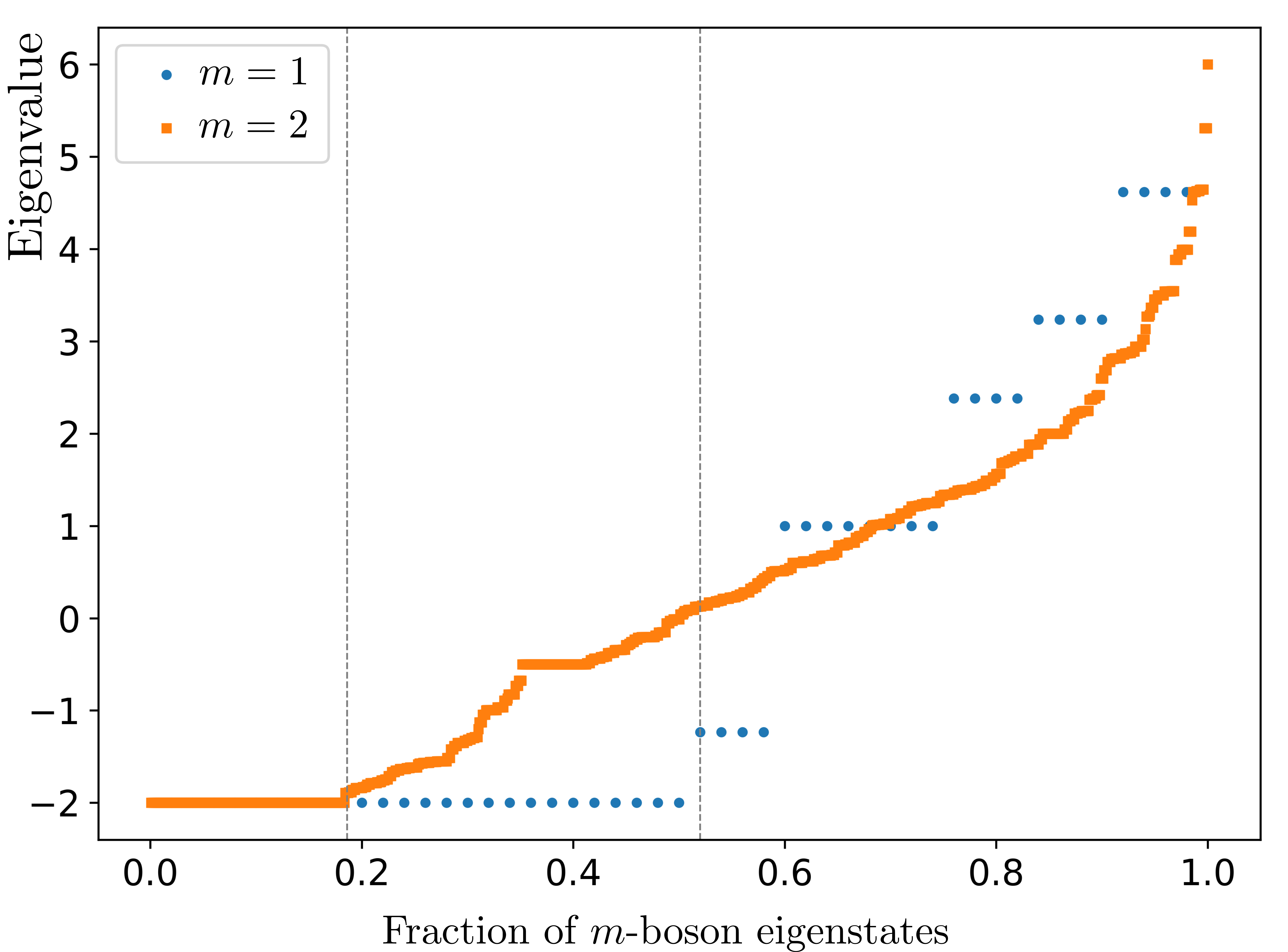}
    \caption{\textbf{Many-body ground state degeneracies in strong-coupling effective Hamiltonian:} Eigenvalues of $m$-boson eigenstates from exact diagonalisation of Eq.~\eqref{eq:boson} for $t_b=1,\mu=0$ on a $5\times 5$ supercell with periodic boundary conditions. The dashed vertical lines indicate the expected fraction of degenerate eigenstates $(N+1)/(2N)$ for $m=1$ and $(N-1)(N-6)/(2N(2N-1))$ for $m=2$. Note that the eigenvalues are shifted by a constant value $3(N-m)t_b/2$ and then normalised by $m$ to highlight the degeneracy from CLS with energy  $-2t_b$ each.}
    \label{fig:ED}
\end{figure}

Next, we consider many-boson states. Clearly, if the bosons occupy CLS on non-overlapping plaquettes, the Hamiltonian in Eq.~\eqref{eq:boson} does not mix the single-boson states, and the bosons remain localised under unitary evolution. For two bosons, there are 5 plaquettes excluded by choosing one plaquette for the first boson, and therefore $(N-1)(N-6)/2$ states that are degenerate with energy $E=-2t_b\times 2$. Extending this line of reasoning to the $\binom{2N}{m}$ states in the $m$-boson Hilbert space for a finite boson-density, we conclude that a fraction $\mathcal{O}(1/2^m)$ of them are part of the degenerate ground state manifold with $E=-2t_b\times m$.  In Fig.~\ref{fig:ED}, we show using exact diagonalisation of Eq.~\eqref{eq:boson} in the $m$-boson sector that the finite ground-state degeneracy of localised eigenstates survives to a finite low density and finite nearest-neighbour repulsion.

\subsection{Superfluid stiffness from exact diagonalisation}\label{subsec:stiffnessED}

\begin{figure}
    \centering
    \includegraphics[width=0.8\linewidth]{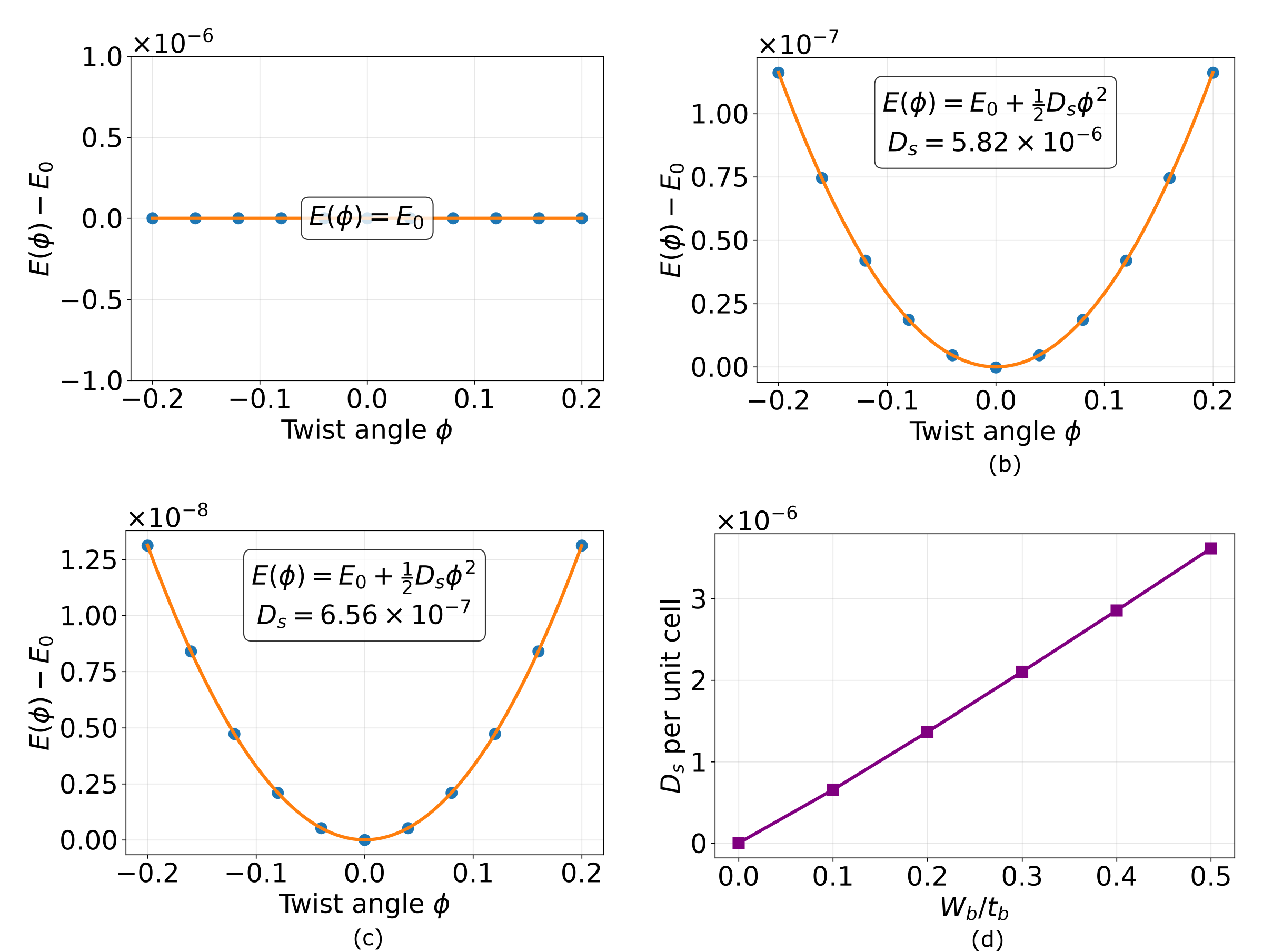}
    \caption{\textbf{Superfluid stiffness from the curvature of ground state energy with twisted boundary conditions:}
    Ground state energy from exact diagonalisation in the presence of a twist $\phi$ in the boundary condition $t_{i\alpha, j\beta}=t_{i\alpha,j\beta}e^{i\phi}$ if the unit cells  $i,j$ are on the x-boundary of the system. The yellow curve is a fit to a quadratic function of $\phi$, $E(\phi) = A_0 + \frac{1}{2} D_s \phi^2$, with the superfluid stiffness $D_s$~\cite{scalapino1993}.
    (a) 4 hard-core bosons on a $5\times 4$ checkerboard lattice ($1/10$ filling) with the narrow band width $W_b=0$
    (b) 8 hard-core bosons on a $4\times 4$ lattice ($1/4$ filling) with $W_b=0$
    (c) 4 bosons on a $5\times 4$ lattice ($1/10$ filling) with $W_b=0.1t_b$
    (d) Superfluid stiffness vs $W_b/t_b$ for 4 bosons on a $5\times 4$ lattice ($1/10$ filling). 
    }
    \label{fig:EDstiffness}
\end{figure}

The zero temperature superfluid stiffness can be derived from the dependence of the ground state energy on the magnitude of a twist in the $U(1)$ phase applied to the boundary conditions~\cite{scalapino1993}. We calculate the ground state energy as a function of twist $\phi$ in the boson hopping integrals crossing the x boundary, $t_{i\alpha, j\beta}=t_{i\alpha,j\beta}e^{i\phi}$ and show the dependence of the superfluid stiffness at low-density as a function of the narrow band width $W_b$ in Fig~\ref{fig:EDstiffness}. 

Consistent with the intuition from Sec.~\ref{subsec:MB}, we find in panel (a) no stiffness in the low-density regime where hard-core bosons are able to occupy non-overlapping plaquettes, with an extensive ground state degeneracy. In contrast, we see in panel (b) that when the density is higher, there is a finite stiffness even in the flat band limit $W_b=0$ due to the finite interactions between pairs. Consistent with the intuition from Sec.~\ref{sec:Hubbard}, in the low-density limit, we find in panel (d) that the stiffness in the strong-coupling expansion truncated to leading order vanishes as we approach the linegraph condition $W\to0 \implies W_b\to 0$ corresponding to exactly flat bands.

\subsection{Vanishing upper bound on superfluid stiffness}

For a small finite $W>0$, the natural many-body ground state for the boson Hamiltonian in Eq.~\eqref{eq:boson} is a condensate at the band minimum at $\Gamma$ - a uniform $d$-wave superfluid. We now present a heuristic upper bound on the superfluid stiffness of this ground state and show that it vanishes in the limit $W\to0$. 

The stiffness is given by the usual transverse limit of the current response to a static vector potential ${\bf A}=A_x\hat{x}$, $D_s=\widetilde{D}-\chi_{jj}(q_\perp\to0,\omega=0)$~\cite{scalapino1993}.  
Since the paramagnetic current response $\chi_{jj}$ is positive-definite, the diamagnetic response $\widetilde{D}(T)$ is a rigorous upper bound on the stiffness $D_s(T)$ at any temperature $T$~\cite{hazra2019}.
\begin{align}
D_{s} & \le\widetilde{D}=\left\langle \partial_{A_{x}}^{2}H_{{\rm hop}}\right\rangle =\sum_{{\bf k}}\left\langle B_{{\bf k}}^{\dagger}\partial_{k_{x}}^{2}h_{{\bf k}}B_{{\bf k}}\right\rangle .
\end{align}
Note that this is the diamagnetic response for any interaction strength. Since the nearest-neighbour repulsion and the infinite hard-core repulsion do not couple to the vector potential $\mathbf{A}$, they can only affect the correlation function $\langle B_{{\bf k}}^{\dagger}B_{{\bf k}} \rangle$.
We can decompose the sum over momenta into an extensive contribution from the condensate at ${\bf k} = \Gamma$ and a contribution from the tail of the boson density distribution function
\begin{align}
    \widetilde{D}=\left\langle B_{\Gamma}^{\dagger}\partial_{k_{x}}^{2}h_{{\bf k}=\Gamma}B_{\Gamma}\right\rangle +\sum_{{\bf k}\neq \Gamma}\left\langle B_{{\bf k}}^{\dagger}\partial_{k_{x}}^{2}h_{{\bf k}}B_{{\bf k}}\right\rangle \label{eq:separation}
\end{align}
which is assumed to be intensive in the limit of low-density of holes. Then $\bfk=\Gamma$ dominates the sum, $\widetilde{D}\approx M^{-1}_\Gamma n_0$ where the matrix element is the expectation value of the inverse mass tensor $M^{-1}_\Gamma = \left\langle \Psi_{\Gamma{\rm flat}}|\partial_{k_{x}}^{2}h_{{\bf k}=\Gamma}|\Psi_{\Gamma{\rm flat}}\right\rangle $ and $n_0$ is the condensate density.
Observing that the lower band eigenstate at $\Gamma$ is exactly $\left(1,-1\right)/\sqrt{2}$  and 
\begin{equation}
\partial_{k_{x}}^{2}h_{{\bf k}=\Gamma}=-t_{b}\left(
\begin{array}{cc}
0 & \cos\frac{k_{x}}{2}\cos\frac{k_{y}}{2}\\
\cos\frac{k_{x}}{2}\cos\frac{k_{y}}{2} & (2-\frac{W_b}{2t_b})\cos k_{x}
\end{array}
\right),\nonumber
\end{equation}
we evaluate the matrix element $M^{-1}_\Gamma=
\left(t_{b}-W_b/4\right)\left(\langle\tau_{z}\rangle-1\right)-t_{b}\langle\tau_{x}\rangle= W_b/4$, where $\tau_{i}$ are Pauli matrices in sublattice (link-orbital)
space, and find that it vanishes as $W\to0^{+}$~\footnote{The dependence on $W$ is not analytic, because the band minimum is not at $\Gamma$ for $W<0$}. Note that the matrix element depends only on the kinetic part of the Hamiltonian $H_{\rm hop}$ and the interactions between bosons affect only the condensate density and the tail of the density distribution, not the vanishing pre-factor. Independent of the strength of nearest-neighbour repulsion between pairs in Eq.~\eqref{eq:boson}, the upper-bound on superfluid stiffness vanishes because the matrix element $M^{-1}_\Gamma\to 0$ in the flat-band limit. Thus, in this low-density, vanishing bandwidth limit, we have 
$d$-wave pairs with no superfluid stiffness.~\footnote{Note that unlike the rigorous upper bounds on superfluid stiffness
in Ref.$\ $\cite{hazra2019,verma2021}, that cannot be violated,
this is a heuristic upper bound derived by neglecting positive-definite
contributions from ${\bf k}\neq\Gamma$ in Eq.~\ref{eq:separation}. It can be exceeded if the
Tan tails in the density distribution $n_{{\bf k}}^{b}=\left\langle B_{{\bf k}}^{\dagger}B_{{\bf k}}\right\rangle$
decay much slower than the $1/k^{4}$ that is characteristic of contact
interactions~\cite{tan2008,tan2008a}. If $n_{{\bf k}}^{b}\propto1/k^{2}$ or slower, then the
contributions from ${\bf k}\neq\Gamma$ in Eq.~\ref{eq:separation} are extensive and comparable to
the ${\bf k}=\Gamma$ contribution.} 

Although the results above rely entirely on the frustration inherent in the pair-hopping Hamiltonian $H_{\rm hop}$, it is clear that both key results: the many-body ground state degeneracy and the vanishing stiffness survive an arbitrary strength of nearest neighbour repulsion between pairs. 
Furthermore, it is crucial that the pair-hopping integral is positive, a negative pair-hopping integral~\cite{hirsch1987} favours $s$-wave pairing and none of the physics of vanishing stiffness and localised pairs in CLS survives in this regime.

\subsection{Implications for lack of magnetic order in spin-1/2 models}

Since the Hilbert space of hard-core bosons is isomorphic to that of spin-1/2 degrees of freedom, the lack of superfluidity at low doping implies a lack of XY order for nearest-neighbour Heisenberg Hamiltonians on the checkerboard line-graph, close to the $S_z$ spin-polarised limit. Specifically, under the mapping $S^+\to b^\dagger, S^-\to b, S_z\to b^\dagger b -1/2$~\cite{matsubara1956}, the pair-hopping term maps to XY spin-exchange and the repulsion maps to Ising spin-interaction, so that the low-energy effective Hamiltonian of Eq.~\eqref{eq:boson} maps to the XXZ Hamiltonian
\begin{align}
    H=\sum_{\langle ij\rangle} J_\parallel (S_i^x S_j^x + S_i^y S_j^y ) + J_z S_i^z S_j^z \label{eq:XXZ}
\end{align}
with $J_\parallel=J_z = 2 t_b$. 
We will consider the more general case when $J_\parallel\neq J_z$, which corresponds to the boson Hamiltonian \eqref{eq:boson} reproduced below 
\begin{align}
    H=\sum_{\langle ij\rangle} t_b (b_i^\dagger b_j + {\rm h.c.}) + V_b (n_i^b -\frac{1}{2}) (n_j^b -\frac{1}{2}) \notag
\end{align}
with $V_b = J_z \neq 2t_b$.
The superfluid stiffness of \eqref{eq:boson} then maps to the spin-stiffness~\cite{kopietzSpinConductanceDynamic1998} of \eqref{eq:XXZ}, the response of the $S_z$ spin-current to a long-wavelength twist of the spins ${\bf S}_j\to e^{iS_z(A_z x_j)}{\bf S}_j e^{-iS_z(A_z x_j)}$.  Under this twist, the Heisenberg Hamiltonian transforms as 
\begin{align}
        H&=\sum_{\langle ij\rangle} \frac{J_\parallel}{2} (S_i^+ S_j^-e^{i\hbar A_z (x_i -x_j)} +{\rm h.c.}) + J_z S_i^z S_j^z  
\end{align}
from which we see that the Ising component of the exchange does not couple to the spin-twist. Thus the vanishing bound on the spin-stiffness holds even away from the Heisenberg point when $J_\parallel\neq J_z$ and implies that the ground state does not break the $U(1)$ spin rotation symmetry in a nearly $S_z$-polarised spin system on the checkerboard line-graph. Intuitively, this is because flipped spins occupy compact localised states, resulting in a ground state manifold comprising a fraction $1/2^m$ of the states in the Hilbert space of fixed $S_z$, where $m=N-S_z/\hbar$ is the number of flipped spins. Note again that both the ground state degeneracy and the bound on the stiffness survive an arbitrary strength of Ising spin-exchange, implying a lack of XY spin-order that survives any anisotropy. This is consistent with existing results on the antiferromagnetic Heisenberg model on the checkerboard lattice where there is general consensus that the ground state is a plaquette valence bond solid with quadrumer long-range order and broken space-group symmetries, but no broken spin-rotation symmetry~\cite{starykh2005}. Although the strong-coupling expansion of the attractive Hubbard model in Sec.~\ref{sec:Hubbard} yields a Heisenberg model as the low-energy effective theory only in the case of time-reversal antisymmetric hopping, XXZ anisotropy is readily induced by a general time-reversal breaking fermion hopping.

\section{Insights from exact ground state of a quantum dimer model}

Complementary to the intuition at low density, the absence of superfluid or XY stiffness can be understood from a different conceptual origin in the Ising anisotropy limit ($J_z\gg J_\parallel$)
of Eq.~\eqref{eq:XXZ} 
when a quarter of the sites have flipped spins. This corresponds to the limit of NN repulsion much stronger than boson-hopping amplitude $H\to H_V=V_{b}\sum_{\left\langle i\alpha,j\beta\right\rangle }n_{i\alpha}^{b}n_{j\beta}^{b}$, with $n_{i\alpha}^{b}$$=b^\dagger_{i\alpha} b_{i\alpha}$ where each boson placed on a link imposes a constraint of no-occupancy on the six neighbouring links on the checkerboard line-graph. The local Hilbert space of a hard-core boson on the line-graph is then isomorphic to that of a dimer covering the occupied link of the underlying (square) lattice,
with the usual dimer constraint that no more than one dimer is incident on every site\footnote{This local equivalence does not hold for the repulsion of particle-hole symmetric densities $\tilde{n}_{i\alpha}^{b}=(b^\dagger_{i\alpha} b_{i\alpha}-1/2)$, but at fixed filling, the difference between these two Hamiltonians $\sum_{\left\langle i\alpha,j\beta\right\rangle }n_{i\alpha}^{b}n_{j\beta}^{b}$ and $\sum_{\left\langle i\alpha,j\beta\right\rangle }\tilde{n}_{i\alpha}^{b}\tilde{n}_{j\beta}^{b}$ is a constant that is dropped in the subsequent analysis without loss of generality.}. 
Quarter filling of hard-core bosons on the line-graph realises the close packing condition of dimers. Then, at quarter-filling, there is no phase-space for single-boson-hopping - all configurations reached by a single boson-hop violate the dimer constraint. There is a macroscopic degeneracy of all possible dimer coverings. 

\begin{center}
    \includegraphics[width=0.5\textwidth]{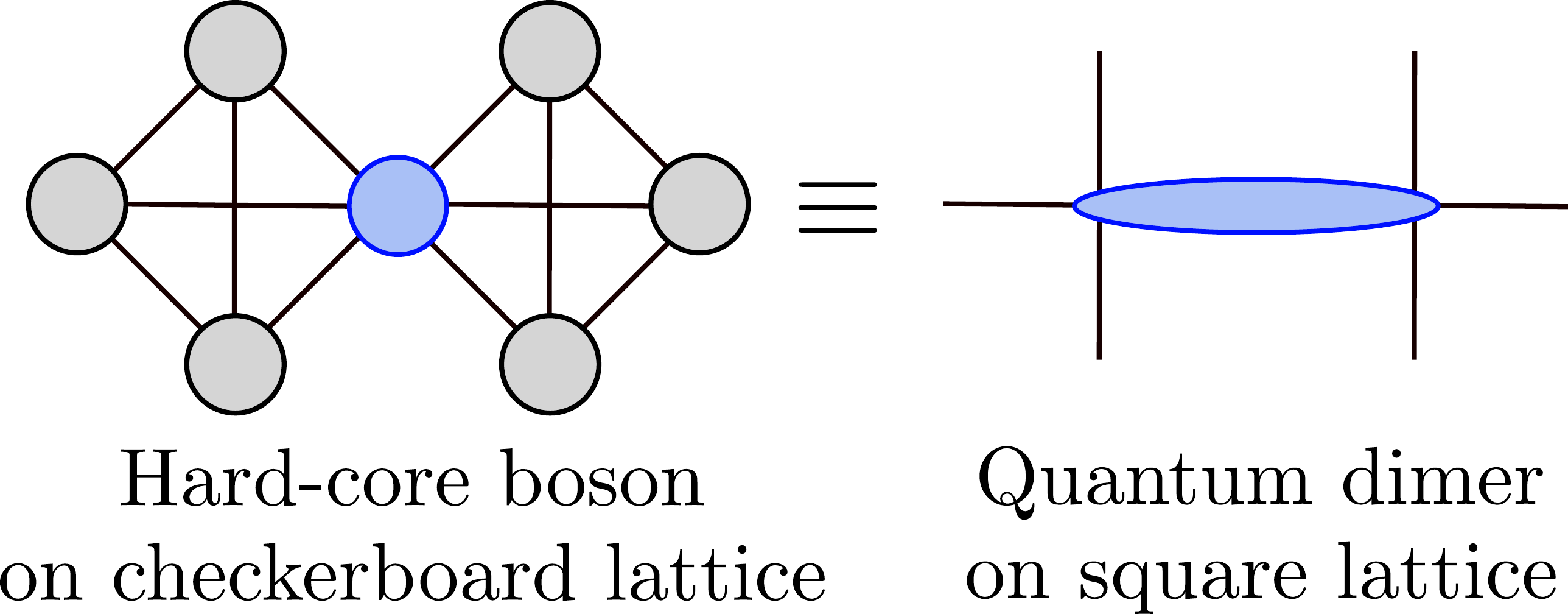}
\end{center}

\subsection{Exact mapping to Rokhsar-Kivelson point on a single plaquette}

Now we consider boson hopping as a perturbation $H=H_V+H_t$ with $H_t = it_{b}\sum_{\left\langle i\alpha,j\beta\right\rangle }\left(b_{i\alpha}^{\dagger}b_{j\beta}+{\rm h.c.}\right)$, with the purely imaginary boson-hopping, breaking time reversal invariance, in contrast to the positive real hopping in \eqref{eq:boson} derived earlier~\footnote{The positive real hopping derived in Eq.~\ref{eq:boson} would yield a conventional s-wave RVB spin-liquid ground state in the analysis that follows.}. 
The imaginary hopping implies a bosonic loop current order on the directed checkerboard graph (Fig.~\ref{fig:DiGraph}) with a Haldane phase of $\pi/2$, which maximises the current for fixed $t_b$.
    At second order in perturbation theory (See Appendix~\ref{app:SW2} for details), there 
is an emergent kinetic term that resonates between two valence bond coverings $\hdimer$, $\vdimer$ of a square plaquette with two dimers on the links. If we consider a single such `flippable' plaquette in isolation, there is a potential energy term of the same magnitude, since both the kinetic and the potential terms emerge through one of four intermediate configurations $\violI$, $\violII$, $\violIII$, $\violIV$ that violate the dimer constraint and each cost energy $V_b$. Thus the low-energy effective Hamiltonian that emerges from a Schrieffer-Wolff transformation in the $t_b\ll V_b$ limit is a quantum dimer Hamiltonian at the Rokhsar-Kivelson point 

\begin{equation}
H = \frac{4t_b^2}{V_b} 
\left(\ket{\hdimer}\bra{\vdimer}+ \text{h.c.}\right)
-  \frac{4t_b^2}{V_b} \left(\ket{\vdimer}\bra{\vdimer} + \ket{\hdimer}\bra{\hdimer} \right)\label{eq:RK1}
\end{equation}

\begin{center}
    \includegraphics[width=0.5\textwidth]{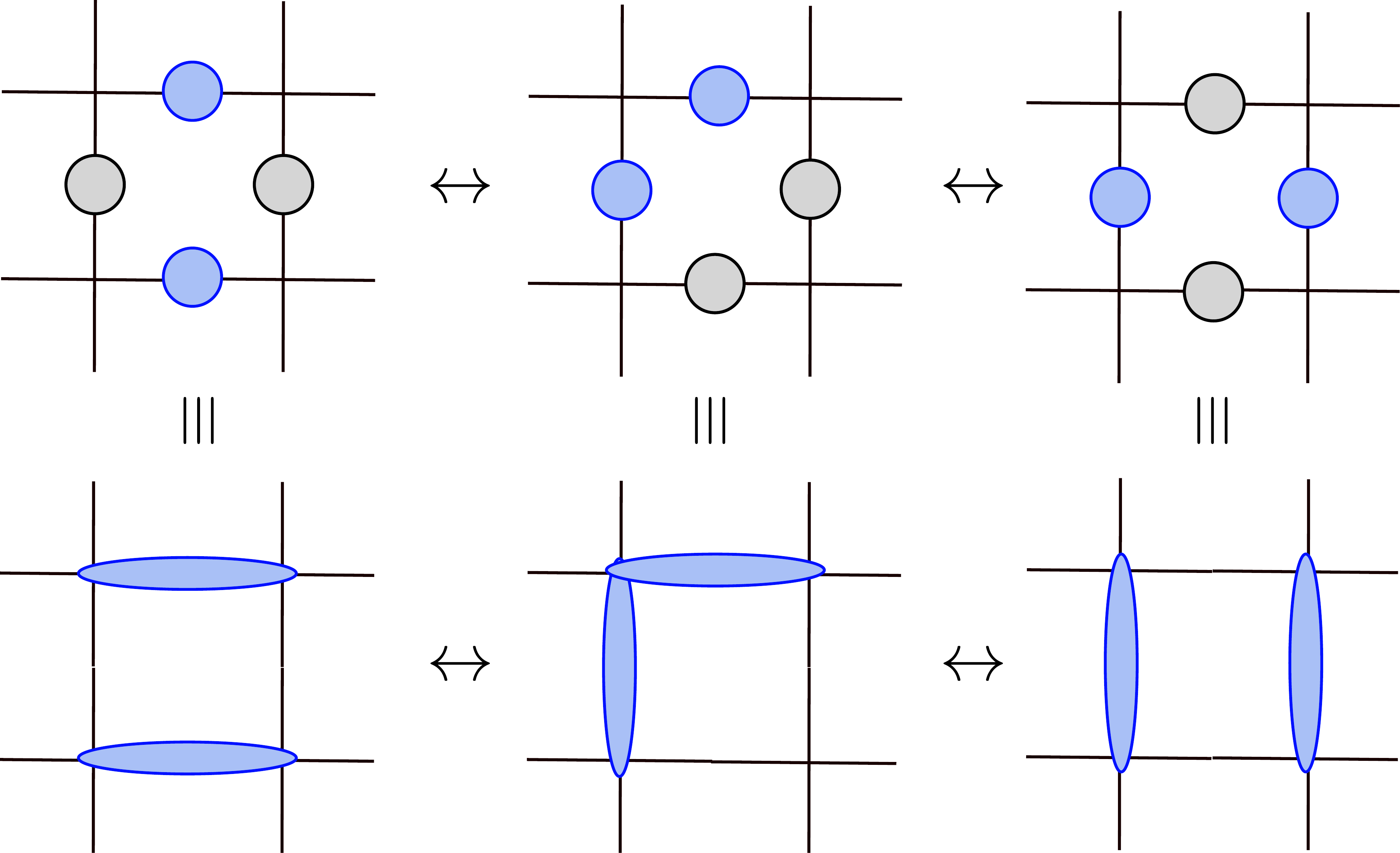}
\end{center}
 
This is the negative of the usual Rokhsar-Kivelson Hamiltonian. 
The ground state is therefore a $d$-wave RVB state, a superposition of the two bosons/dimers occupying the horizontal/vertical links with a relative minus sign $\ket{\Psi_{GS}} = \left( \ket{\hdimer} - \ket{\vdimer} \right)/\sqrt{2}$. The usual equal-superposition ground state $\left( \ket{\hdimer} + \ket{\vdimer} \right)/\sqrt{2}$ is an excitation with $E=8t_b^2/V_b$.

\subsection{Adiabatic deformation to Rokhsar-Kivelson point on a finite lattice}

Next we consider a $2M\times 2N$ square lattice, with periodic boundaries and close-packed dimers. Here the potential energy that is diagonal in the dimer coverings appears on all plaquettes, flippable or otherwise, so that the low-energy effective Hamiltonian is not an emergent Rokhsar Kivelson model. In fact, since each dimer can resonate to the six links adjacent to it, the potential energy is simply $-6t_b^2/V_b\times N_b$, where $N_b=2MN$ is the number of dimers or bosons on the lattice, which is a constant that we drop in the subsequent discussion.

\begin{center}
    \includegraphics[width=0.2\linewidth]{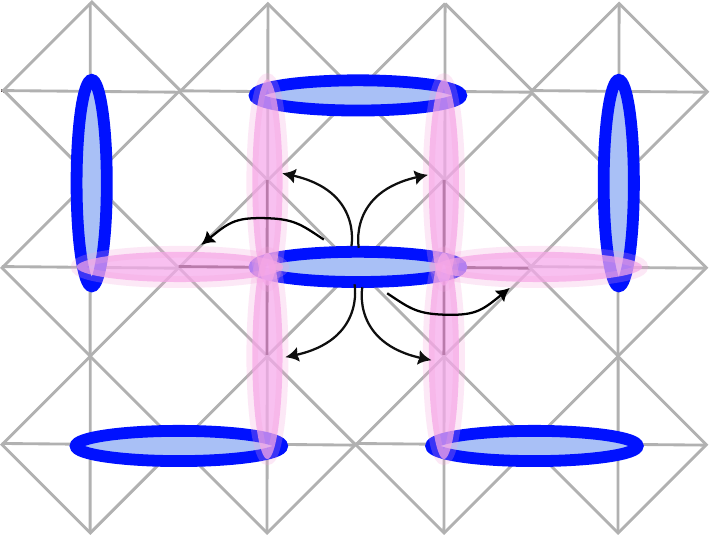}
\end{center}

The low-energy effective Hamiltonian in the Ising limit on a finite lattice is then
\begin{align}
    H_{\rm flip} = \frac{4t_b^2}{V_b} \sum_{\square}^\prime 
    \left(\ket{\hdimer}\bra{\vdimer}+ \text{h.c.}\right)
    \label{eq:Hflip}
\end{align}
where the sum is over all flippable plaquettes. We will find that the ground state is smoothly connected to the ground state of the Rokhsar-Kivelson Hamiltonian
\begin{equation}
H_{\rm RK} = \frac{4t_b^2}{V_b} \sum_{\square}^\prime
\left(\ket{\hdimer}\bra{\vdimer}+ \text{h.c.}\right)
+ \left(\ket{\vdimer}\bra{\vdimer} + \ket{\hdimer}\bra{\hdimer} \right)\label{eq:RK}
\end{equation}
which imposes an additional energy cost of $4t_b^2/V_b$ for each flippable plaquette in a dimer covering.

\begin{figure}
    \centering
    \includegraphics[width=\linewidth]{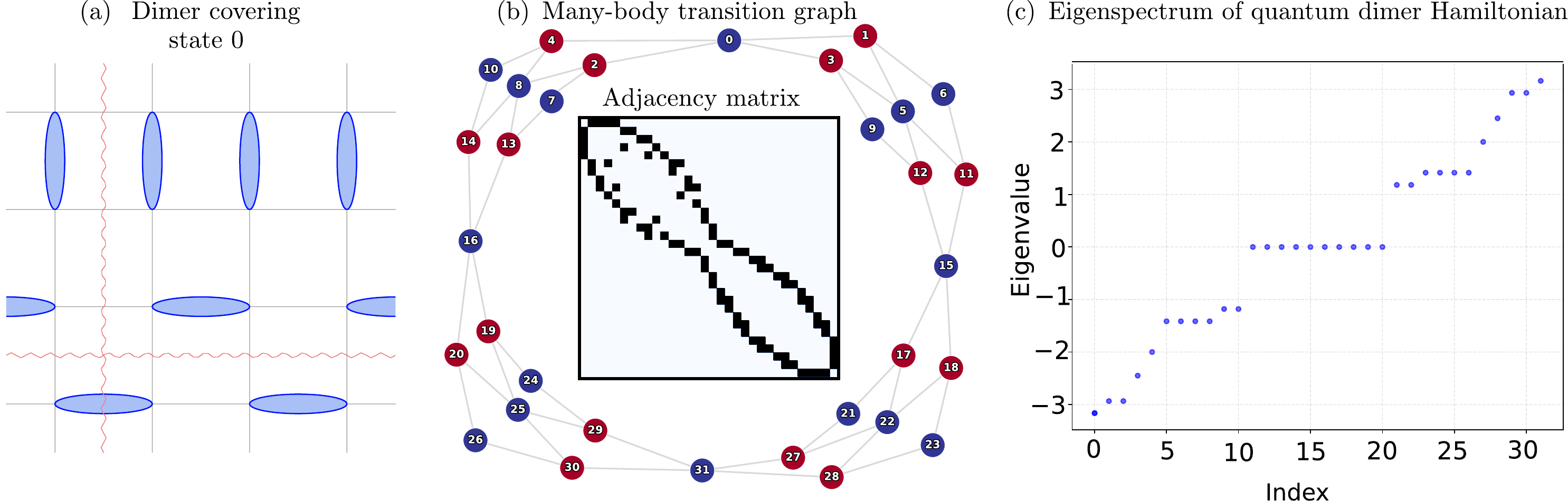}
    \caption{\textbf{Exact diagonalisation in the topological sector of the quantum dimer model}: Since plaquette flips can only change the number of dimers ($N_h,N_v$) crossing any horizontal or vertical cuts (red) through the periodic lattice by units of 2, the odd/even parity of $N_h,N_v$ defines four topological super-selection sectors that are mutually inaccessible by arbitrary plaquette-flips~\cite{moessner2011}. Starting from the dimer covering in (a) for a 4x4 lattice, $H_{\rm flip}$ in Eq.~\eqref{eq:Hflip} induces transitions to the 32 dimer coverings depicted in the state graph in (b). Diagonalising the adjacency matrix of this graph yields the many-body eigenstates of $H_{\rm flip}$ shown in (c). The bipartite nature of the transition graph leads to a particle-hole symmetric spectrum - a $Z_2$ gauge transform on the red nodes (dimer coverings) $|c_{\rm red}\rangle\to -|c_{\rm red}\rangle$ amounts to changing the sign of the dimer resonance energy $4t_b^/V_b \to -4t_b^2/V_b$ and effectively inverting the spectrum. Note the presence of a degenerate set of zero-modes in (c) that we will discuss later.}
    \label{fig:dimerdiagonalization}
\end{figure}

\begin{figure}
    \centering
    \includegraphics[width=\linewidth]{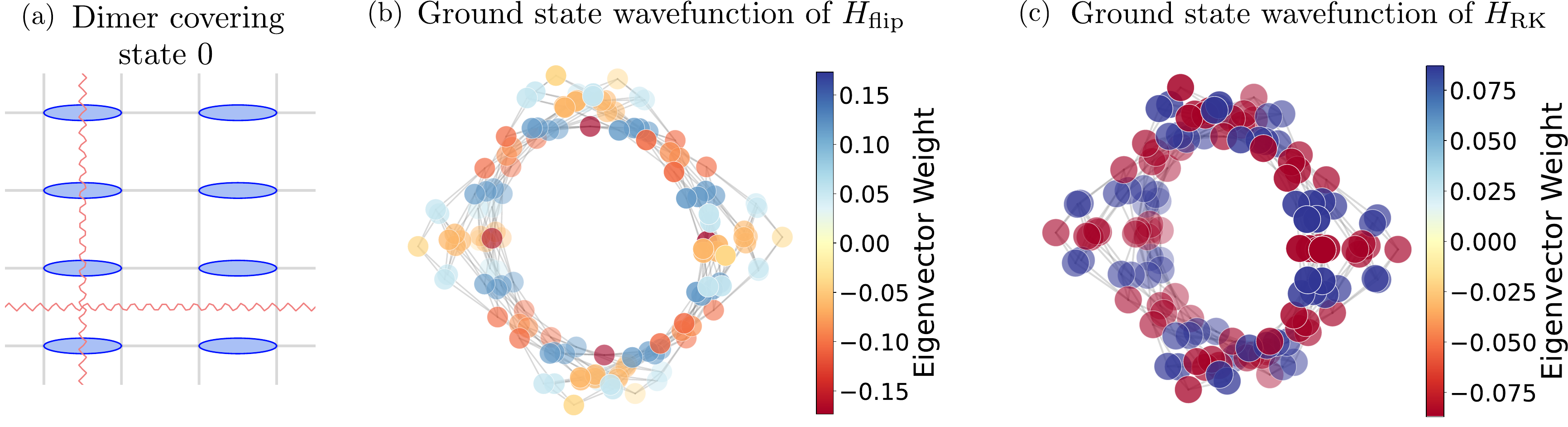}
    \caption{\textbf{Exact diagonalisation in the non-topological sector of the quantum dimer model}: Starting from the `columnar' dimer covering (a) in the topologically trivial sector with $(N_h,N_v)=$ (even,even), the Hilbert space of 132 dimer coverings spanned by plaquette-flips on a 4$\times$4 lattice is connected by the transition graph shown in (b) and (c), on which the ground state wavefunctions of $H_{\rm flip}$ and $H_{\rm RK}$ are shown in colour-scale.}
    \label{fig:dimergs}
\end{figure}

We diagonalise these two Hamiltonians and find their eigenspectra as described in Fig.~\ref{fig:dimerdiagonalization},\ref{fig:dimergs}. The ground state of the Rokhsar-Kivelson Hamiltonian $H_{\rm RK}$ is known exactly on any finite-sized lattice to be 
\[|\Psi_{0,RK}\rangle \equiv N_c^{-1/2} \sum (-1)^F  | c \rangle \]
where $|c\rangle$ is a chosen reference dimer covering, the sum is over all dimer coverings that are accessible from $| c \rangle $ by a sequence of plaquette flips~\cite{moessner2011,ardonne2004}, $F$ is the length of this sequence, and $N_c$ is the number of such dimer coverings. This is because the Hamiltonian $H_{\rm RK}$ is a positive semi-definite operator which can be written as a sum of quadrature terms
\begin{equation}
    H_{\rm RK} = \frac{4t_b^2}{V_b} \sum_{\square}^\prime
    \left( \ket{\hdimer} + \ket{\vdimer} \right)
    \left( \bra{\hdimer} + \bra{\vdimer} \right)
\end{equation}
on each flippable plaquette~\cite{
kivelsonTopologyResonatingValencebond1987,rokhsar1988} so that the expectation value $\langle \Psi|H|\Psi\rangle\ge 0$ for any arbitrary state $|\Psi\rangle$. Since the eigenspectrum is lower-bounded by zero, a state that is annihilated 
by $\left( \bra{\hdimer} + \bra{\vdimer} \right)$ on each flippable plaquette is an exact ground state. One can readily check that $|\Psi_{0,RK}\rangle$ has this property. In Fig.~\ref{fig:dimergs}(c), we reproduce this wavefunction $|\Psi_{0,RK}\rangle$ on a $4\times4$ lattice, which has equal weight on all nodes (dimer coverings) with an alternating sign on adjacent nodes that are connected by a plaquette flip. The ground state of $H_{\rm flip}$ (Fig.~\ref{fig:dimergs}(b)) is also a coherent superposition of all dimer coverings with the same alternating sign-structure. However, it is not an equal-weight superposition, with the maximum weight on the `hubs' of the many-body transition graph that correspond to the so-called `columnar' configurations with the maximum number of flippable plaquettes.

\begin{figure}
    \centering
    \includegraphics[width=\linewidth]{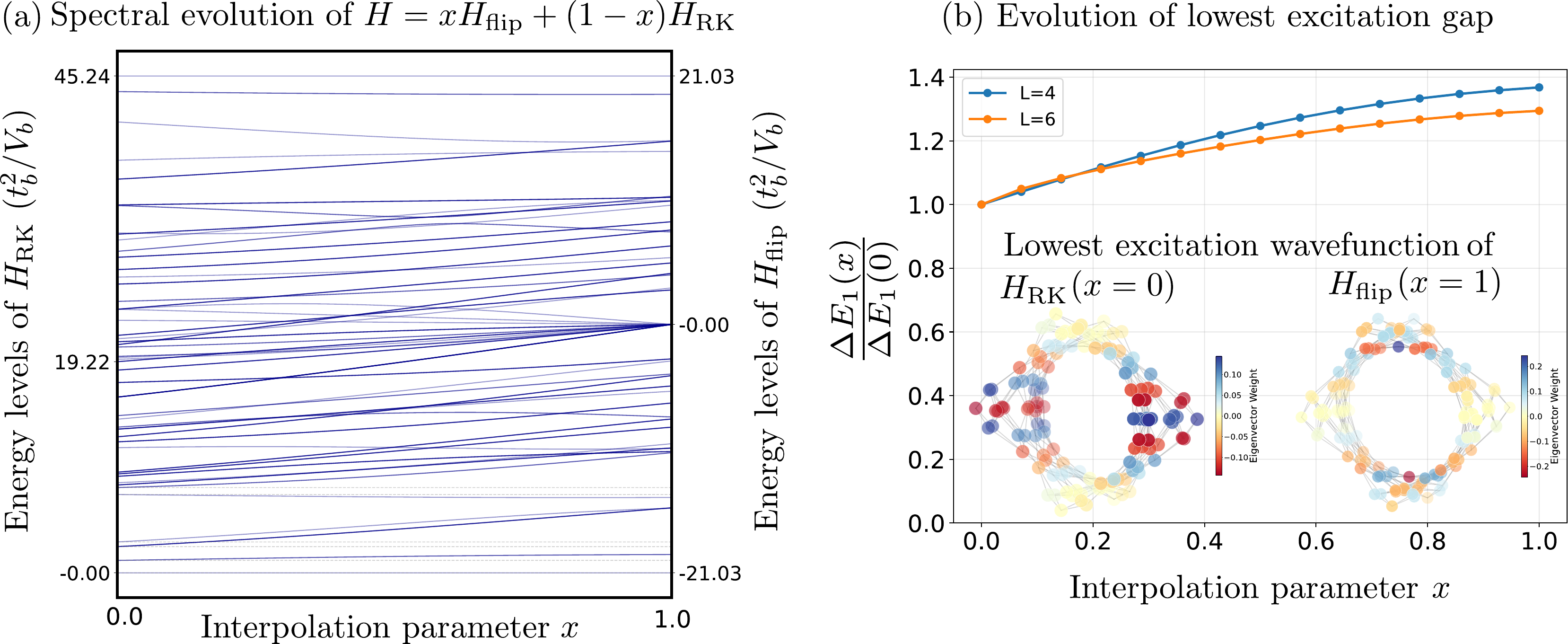}
    \caption{\textbf{Interpolation of excitation energies from $H_{\rm RK}$ to $H_{\rm flip}$:} (a) Evolution of energy levels of $H_{\rm RK}$ at $x=0$ to those of $H_{\rm flip}$ at $x=1$ on a $4\times4$ lattice in the topologically trivial sector (Fig.~\ref{fig:dimergs}(a)). (b) The finite-size gap to the lowest excitation grows as we evolve from $H_{\rm RK}$ to $H_{\rm flip}$ for $4\times4$ and for $6\times6$ lattices. These excitations are doubly degenerate and correspond to two orthogonal standing waves ($\approx$ $\cos$ and $\sin$) on the cyclic transition graph or insertions of $\pm 2\pi$ flux through the hole in the graph. These gapless excitations are qualitatively similar for the Hamiltonians $H_{\rm RK}$ and $H_{\rm flip}$.}
    \label{fig:dimerspectrum}
\end{figure}

To establish that the ground states of these two Hamiltonians are smoothly connected, we follow the energy gap to the lowest lying excitations as we interpolate from the Rokhsar-Kivelson Hamiltonian $H_{\rm RK}$ to the purely kinetic Hamiltonian $H_{\rm flip}$ (see Fig.~\ref{fig:dimerspectrum}). The low energy excitations of the quantum dimer models in the close-packed sector are so-called gapless `resonon' and `$\pi$0n' modes~\cite{rokhsar1988,moessner2003a,lauchli2008}, that amount to threading a flux through the (cyclic) many-body transition graph. The wavefunctions of the lowest energy modes are shown in the inset of Fig.~\ref{fig:dimerspectrum}(b). On finite-sized lattices, we find the gap to these excitations grows as we interpolate from the exactly-solvable $H_{\rm RK}$
to the kinetic Hamiltonian $H_{\rm flip}$, so that the ground states are smoothly connected by adiabatically turning off the potential energy cost of flippable plaquettes. In Fig.~\ref{fig:dimerspectrum}
(a), we show that the excitation gap to the rest of the spectrum also does not close in this interpolation from $H_{\rm RK}$ to $H_{\rm flip}$.

\subsection{Compact localised many-body eigenstates and extensive ground state degeneracy}
\begin{figure}
    \centering
    \includegraphics[width=0.75\linewidth]{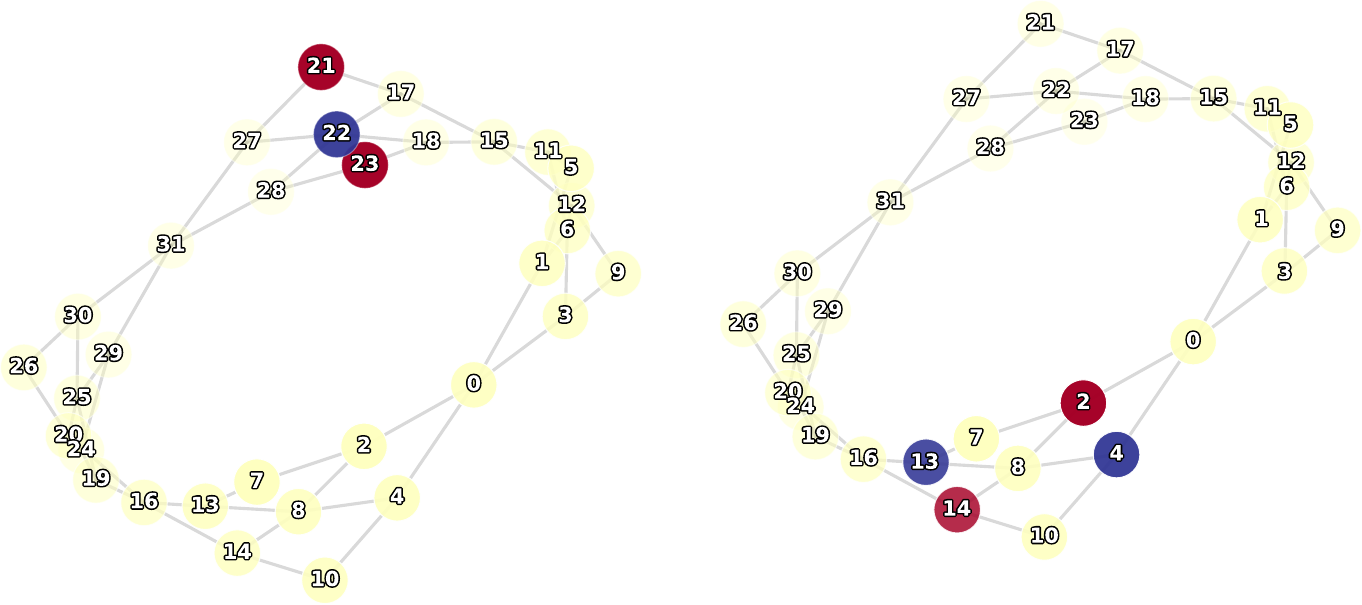}
    \caption{\textbf{Compact localised many-body eigenstates} of $H_{\rm flip}$ in the Hilbert space of a 4x4 lattice in the topologically non-trivial sector with $N_h,N_v=$even, odd (see Fig.~\ref{fig:dimerdiagonalization}). Red and blue colours on the nodes indicate equal positive and negative weight on the corresponding dimer coverings, leading to destructive interference of many-body transitions to adjacent dimer configurations and disorder-free localisation in many-body Hilbert space. Each of these CLS generates an extensive set of degenerate zero-modes by translation along the cyclic transition graph. Note that only the second set of CLS shown here, supported on dimer covering with equal number of flippable plaquettes, remains a compact localised eigenstate of $H_{\rm RK}$ with energy $3\times 4t_b^2/V_b$.}
    \label{fig:dimercls}
\end{figure}

We note that in Fig.~\ref{fig:dimerdiagonalization}, a large fraction of the many-body eigenstates of $H_{\rm flip}$ are degenerate with zero energy. Such many-body degeneracies have been discussed as `many-body flat bands' or `Fock-space cages' or `interference-caged quantum many-body scars' in recent literature~\cite{tan2025,ben-ami2025,jonay2026,mohapatra2026a} and originate from destructive interference of transition pathways on the many-body state graph\footnote{Note also that the dimer Hamiltonian $H_{\rm flip}$ in Eq.~\eqref{eq:Hflip} has an emergent low-energy symmetry corresponding to dipole conservation, connecting to recent discussions of localisation in many-body Hilbert-space~\cite{sala2020}.}. Similar to the non-interacting flat-band states on Lieb-type lattices~\cite{lieb1989}, the zero-energy degenerate eigenstates of $H_{\rm flip}$ can be understood as originating from compact localised states in the transition graph. Specifically, the transition graph in Fig.~\ref{fig:dimerdiagonalization}(b) can be viewed as a periodic decorated-diamond lattice in 1D with the following unit cell
\begin{tikzpicture}[
    every node/.style={circle, draw, minimum size=1.5mm, inner sep=0pt}
]

\node (0) at (0,0) {};
\node (2) at (0.5,0.5) {};
\node (4) at (0.5,-0.5) {};
\node (7) at (1,1) {};
\node (8) at (1,0) {};
\node (10) at (1,-1) {};
\node (13) at (1.5,0.5) {};
\node (14) at (1.5,-0.5) {};
\node (16) at (2,0) {};

\draw (0) -- (2);
\draw (0) -- (4);
\draw (2) -- (7);
\draw (2) -- (8);
\draw (4) -- (8);
\draw (4) -- (10);
\draw (7) -- (13);
\draw (8) -- (13);
\draw (8) -- (14);
\draw (10) -- (14);
\draw (16) -- (13);
\draw (16) -- (14);
\end{tikzpicture}
and hosts two distinct types of CLS in each unit cell as shown in Fig.~\ref{fig:dimercls}.

\begin{figure}
    \centering
    \includegraphics[width=0.8\linewidth]{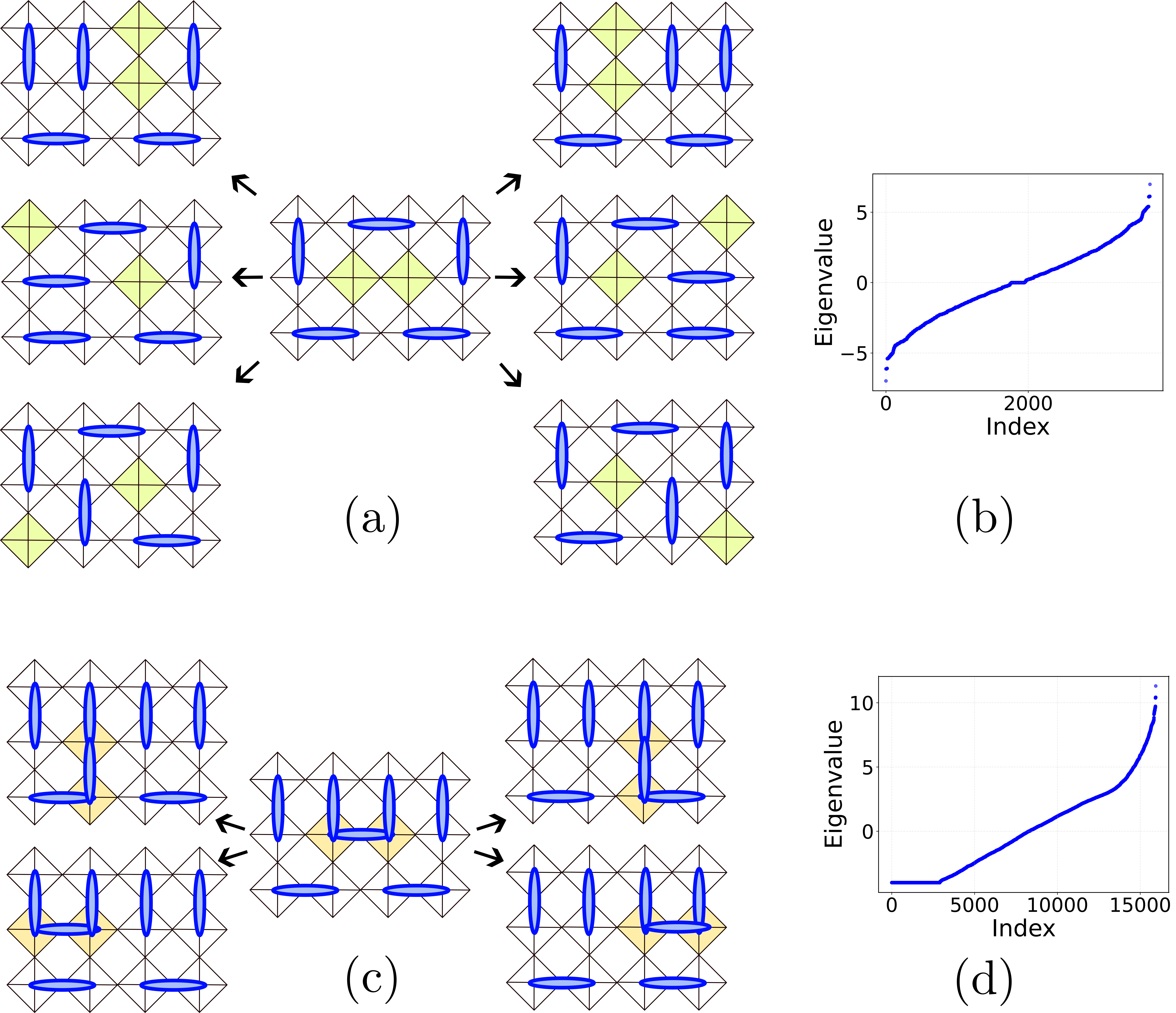}
    \caption{\textbf{Doped hole-pair and electron-pair states above quarter-filling:} (a) Deconfined holons: Six configurations accessible by a single dimer-hop from the central configuration of a single dimer-hole. The two stars which contribute $3V_b/2$ to the Hamiltonian $H_V$ are highlighted in green. These stars (defect sites) with local charge and energy deviation are deconfined with no energy cost to $\mathcal{O}(V_b)$. (c) Extensively degenerate many-body ground states: Four configurations accessible from the central configuration of one dimer in excess of the dimer constraint. The two stars (defect sites) which contribute $-V_b$ to the Hamiltonian are highlighted in yellow. In contrast to the dimer-hole in (a), the excess dimer in (b) is not fractionalised into deconfined partons. The energy of the leading order boson-hopping Hamiltonian $\mathbb{P}H_t\mathbb{P}$ is shown in (b) and (d) in units of $t_b$ for the single hole-pair doped and electron-pair doped sector, showing a unique ground state for the deconfined holons and a large degenerate ground-state manifold for the confined excess dimer states. The projection operator $\mathbb{P}$ excludes dimer moves that change the total number of defect sites. }
    \label{fig:dimerdoped}
\end{figure}

Such extensive degeneracies spanning a finite fraction of total eigenstates also appear in the ground state of a single boson in excess of quarter-filling (ie. a charge-2e excitation of the correlated insulator) as shown in Fig.~\ref{fig:dimerdoped}(d). Introducing this extra boson now allows single-boson hopping processes (Fig.~\ref{fig:dimerdoped}(c)) that do not violate the dimer constraint. The leading order Hamiltonian $\mathbb{P}H_t\mathbb{P} = \mathbb{P}t_{b}\sum_{\left\langle i\alpha,j\beta\right\rangle }\left(b_{i\alpha}^{\dagger}b_{j\beta}+{\rm h.c.}\right)\mathbb{P}$ has the eigenspectrum shown in Fig.~\ref{fig:dimerdoped} with a large degeneracy at low energy that is lifted by the plaquette-flip dynamics at $\mathcal{O}(t_b^2/V_b)$. Here the projection operator $\mathbb{P}$ ensures that the single-dimer hopping does not change the number of defect sites. The implications of this extensive degeneracy on the transport, optical and superfluid response of the doped d-RVB spin-liquid remain to be explored.

\subsection{Deconfined holon excitations}
The ground state of the hard-core boson Hamiltonian at quarter-filling in the Ising limit with time-reversal invariance strongly broken is then a $d$-wave RVB spin-liquid\footnote{Frustration-free Hamiltonians similar to the Rokhsar-Kivelson model have a long history in the context of finding Hamiltonians whose exact ground states have specific superfluid wavefunctions~\cite{arovasExactQuestionsInteresting1992,kane1991,rana1993,han2024a,han2025a} and the absence of superfluid stiffness in our work has no contradiction with this literature}
 state, distinct from the usual short-ranged RVB that is discussed in the context of negative dimer resonance energy, which has no $(-1)^F$ factor in the ($s$-wave) ground state wavefunction\footnote{This is also distinct from the more widely discussed ``vanilla'' d-wave RVB state which is defined by the Gutzwiller projection of a d-wave BCS ground state, and includes resonances of singlets between sites arbitrarily far away~\cite{kotliarResonatingValenceBonds1988,riceDwaveResonanceValence1995,andersonPhysicsHightemperatureSuperconducting2004,liuPairingSuperconductivityDriven2005}. In the strong-pairing limit, where the pair-size is small enough that only nearest-neighbour pairs are important, the ``vanilla'' $d$-wave RVB is smoothly connected to the short-ranged $d$-wave RVB we discuss.}. To see that the doped charge fractionalises into deconfined holons, it is helpful to rewrite the NN repulsion $H_V=$$2V_b\sum_{i\alpha,j\beta} \tilde{n}_{i\alpha}^b \tilde{n}_{j\beta}^b$$=\sum_{\cstar} e_{\cstar}$ as a sum of terms around a site-centered plaquette (star), with $\tilde{n}_{i\alpha}^{b}=(b^\dagger_{i\alpha} b_{i\alpha}-1/2)$. Each such star term encodes all-to-all repulsion between the 6 pairs formed between the 4 links in the star, and has energy 
$V_b/2,0,-2V_b,0,3V_b$ 
for local configurations with $0,1,2,3,4$ dimers, respectively. At quarter-filling, all stars have one dimer each. Removing one boson (dimer) on the checkerboard (square) lattice creates two stars with no dimers on them, and single-boson hopping creates configurations where these ``charged'' stars are arbitrarily separated with no energy cost to $O(V_b)$. On a bipartite lattice such as the square lattice, the charge on stars on each sublattice is separately conserved, since these partons can only move to the next-neighbour or next-to-next-neighbour sites by a single-boson-hop. Similar to the well-known deconfined excitations of the conventional (s-wave) RVB state, these correspond to charge-e, spin-0 partons of the bosons(dimers) that are deconfined on the underlying lattice (See Fig.~\ref{fig:dimerdoped}), with a unique ground state in the sector with one boson less than quarter-filling.

\section{Discussion}

The strong-coupling limit of flat band superconductivity offers many surprises. Conventional wisdom dictates that on a lattice, the strong-coupling Bose-Einstein condensate limit exhibits a wide separation between pair-binding energies ($\mathcal{O}(U)$) and pair-coherence energy scales ($\mathcal{O}(t^2/U)$). In line-graph lattices with obstructed flat bands in presence of strong time-reversal breaking, this separation is parametrically larger with the pair-coherence energy scaling as $\mathcal{O}(t^4/U^3)$. This extended regime of phase fluctuations and vanishing superfluid stiffness opens the door to various competing phases.

We thus find two complementary origins of vanishing stiffness in a model of Cooper pairs with nearest-neighbour hopping and repulsion. At low density, this is rooted in the pairs occupying non-overlapping compact localised states. At precisely quarter filling, it is rooted in an exactly solvable limit where the ground state maps to a quantum spin liquid.  The absence of superfluid stiffness to the leading order in the strong-coupling limit at arbitrary density of link-orbitals is anchored in these two limiting regimes (see Fig.~\ref{fig:phasediagram}).

Of course, the absence of stiffness highlighted in the three regimes of the bosonic Hamiltonian in Fig.~\ref{fig:phasediagram} relies crucially on the fine-tuned equality of NN and NNN hopping on the checkerboard graph ($W=0$ in Eq.~\eqref{eq:hk}). Nevertheless, these results anchor a limiting case around which the leading order contribution to the stiffness is expected to be small when the flat band has a small bandwidth $W>0$.

\subsection{Experimental signatures}

The relevance of obstructed Cooper pairs to any material is determined
by experimentally observing its defining features. Obstructed pairs
are defined not by their symmetry, but by their unusual localisation,
even in the limit of no potential disorder. 
Experimentally, if the sign structure of the condensate wavefunction
is directly imaged by scanning Josephson interferometry (SJI)~\cite{coleman2020}, the defining feature of obstructed pairs is spectral weight on the
edges of the lattice with the sign-structure of the flat-band wavefunction
(Fig. \ref{fig:checkerboard}(c)). Additionally, coincidence two-photoemission
(2e-ARPES) is a recently developed experimental technique~\cite{zwettler2024}
that probes the two-particle spectral function instead of the one-particle
spectral function (c.f. ARPES). As discussed in Appendix~\ref{app:2eARPES}, this could directly image the flat
pair dispersion that characterises obstructed pairs (Fig. \ref{fig:checkerboard}(b)),
as well as the symmetry character that is responsible for the obstructed pairs in compact localised states (Fig \ref{fig:checkerboard}(a)). Lastly, the distinction between local pairs that
are more likely to be found on x and y bonds which are neighbours and
those that are just as likely to be further away is statistically
quantifiable in the cross-correlation of observed x-bond pairs and
y-bond pairs, as measured spatially using SJI and temporally using
2e-ARPES. Conclusively ruling out these observations falsifies the
hypothesis of obstructed pairs in a material. 

\section{Conclusion}

The mechanism of disorder-free localisation driven by strong-interactions that we discuss has parallels with other origins of emergent inhomogeneity leading to localisation~\cite{anderson1975,trugman1988,hirsch1987,schmalian2000,smithDisorderFreeLocalization2017}, but it is distinct in many respects from all of these~\footnote{It is also distinct from plaquette-based approaches~\cite{altman2002,tsai2006} where superlattice-periodic ``inhomogeneity'' is added at level of the non-interacting Hamiltonian.}. The extensive ground-state degeneracies arising from the perfect localisation of bosons in CLS is unlikely to survive a finite-ranged Coulomb repulsion. An important open question is to identify the eventual ground state as a weak screened Coulomb interaction is turned on.

In this paper, we have discussed obstructed bosons with spin-0 and
charge-2e, called obstructed Cooper pairs. Similar mechanisms lead
to other families of obstructed bosons, such as obstructed spins in
strongly repulsive Hubbard-type models~\cite{hazra2024}. 
In the quest to understand the interplay of localisation and strong correlations,
obstructed bosons provide a new language that we hope will enable
many new narratives.

\section*{Acknowledgements}
T.H. would like to thank Adam Nahum for an insightful discussion that eventually led to the connections to dimer models, Daniel Schultz for a clear analytical derivation of the fraction of degenerate $m$-boson eigenstates from non-overlapping CLS, Piers Coleman for a critical reading of an earlier draft, and Mohit Randeria for an insightful comment that motivated the mean-field calculations early on. We acknowledge the use of the \href{https://quspin.github.io/QuSpin/}{QuSpin} package~\cite{Weinberg2017,weinberg2019a} for calculating the many-body ground state energy under twisted boundary conditions.

\paragraph{Author contributions}
TH: conceptualisation, methodology, software, formal analysis, investigation, writing (original draft), writing (review and editing). NV:  software, validation, investigation, writing (review and editing). JS: validation, writing (review and editing), supervision.

\paragraph{Funding information}
T.H. was supported by the Alexander von Humboldt Foundation. 
N.V. was supported as part of Programmable Quantum Materials, an Energy Frontier Research Center funded by the U.S. Department of
Energy (DOE), Office of Science, Basic Energy Sciences
(BES), under award DE-SC0019443.
J.S. was supported by the German Research Foundation TRR 288- 422213477 ELASTO-Q-MAT, B01 and grant SFI-MPS-NFS-00006741-05 from the Simons Foundation.

\paragraph{Data availability statement}
The data that support the findings of this article are openly available at \footnote{https://github.com/tamaghnahazra/MeanField and https://github.com/tamaghnahazra/ObstructedPairsExactDiag}.

\begin{appendix}
\numberwithin{equation}{section}

\section{Time-reversal breaking fermion hopping}\label{app:TRBhop}

The condition $t_\uparrow t_\downarrow <0$ in the electronic hopping in \eqref{eq:hk} implies that time-reversal invariance is strongly broken because $t_\uparrow \to t_\downarrow^*$ under time-reversal. In fact, the fermion hopping Hamiltonian is antisymmetric under time-reversal if $t_\uparrow = - t_\downarrow^*$. For a general complex-valued hopping amplitude $t_\sigma = |t_\sigma|e^{i\phi_\sigma}$, the condition $t_\uparrow t_\downarrow <0\implies \phi_\uparrow + \phi_\downarrow = \pi$ (mod $2\pi$) can be satisfied for instance by setting $t_\uparrow = t$ and $t_\downarrow = -t$ for real hopping amplitude $t$, which makes the up and down spin dispersions particle-hole conjugates of each other. This is the case explored in the main text. 

\begin{figure}
    \centering
    \includegraphics[width=0.4\linewidth]{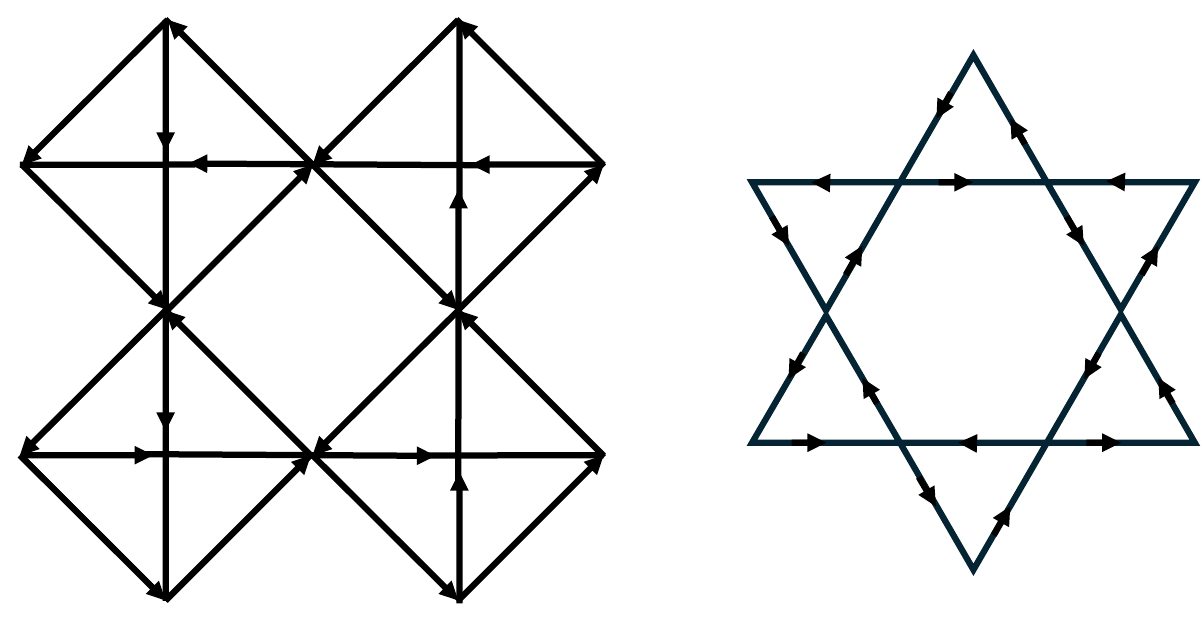}
    \caption{Orientation of the directed checkerboard graph (left) and Kagome graph (right).}
    \label{fig:DiGraph}
\end{figure}

Another possibility is the extreme case of a loop-current order parameter, which can be realised by associating a Haldane phase $\phi$ on the directed subgraph of the edges of the lattice graph on which the hopping integrals are finite, so that the hopping integral on each edge of this subgraph transforms as $t_{ij} \to t_{ij} e^{i\phi}$ if the orientation of the edge $ij$ is from $i$ to $j$ and $t_{ij} \to t_{ij} e^{-i\phi}$ otherwise. On the checkerboard and Kagome line-graphs of the square and honeycomb graphs, a fully time-reversal antisymmetric spin-rotation symmetric electron hopping Hamiltonian is realised by the Haldane phase $\phi=\pi/2$ (which maximises the current for fixed $t$) on the full linegraph with a possible orientation shown in Fig.~\ref{fig:DiGraph}.

The condition $t_\uparrow t_\downarrow < 0$ requires a hopping amplitude whose phase depends on the internal state, and ultracold atoms in optical lattices provide an established toolbox for exactly this kind of engineering. Synthetic gauge fields realised with laser-assisted tunnelling and with periodically driven lattices allow the phase of the tunnelling matrix element to be set at will \cite{dalibard2011, galitski2013}, and all three cases relevant here have been demonstrated. The phase may be the same for all internal states \cite{aidelsburger2011, aidelsburger2013a}, as in the realisation of the Harper Hamiltonian with a gravitational tilt \cite{miyake2013a}. It may be equal and opposite for two spin states, as in the realisation of the Hofstadter Hamiltonian with a magnetic field gradient \cite{aidelsburger2013}, which simultaneously realises the time-reversal invariant Hamiltonian underlying the quantum spin Hall effect; there $t_\uparrow = t_\downarrow^*$ and the pairs remain conventional. Or the two states may differ in phase by $\pi$, which is the sign reversal $t_\downarrow = -t_\uparrow$ required for obstructed pairs; this has been achieved for fermionic $^{40}\mathrm{K}$ by modulating a magnetic field gradient in time, which tunes the relative amplitude and sign of the tunnelling for different internal states \cite{jotzu2015}.

We note that engineering the hopping phase does not by itself enforce the equality of nearest-neighbour and next-nearest-neighbour amplitudes that flattens the band, which remains a separate requirement in any of these schemes.

\section{Superfluid stiffness in mean-field theory}
\label{app:superfluid_stiffness}
Following the approach in \cite{scalapino1993}, we evaluate the superfluid stiffness $D_{s}$ by calculating the change in the Helmholtz free energy under an applied phase twist. The twist is introduced by enforcing twisted boundary conditions on the fermionic fields across a supercell of length $L_x$:
\begin{equation}
\psi_\alpha(\mathbf{r}_i + L_x \mathbf{e}_x) = e^{i\Phi} \psi_\alpha(\mathbf{r}_i) \quad \forall i, \alpha,
\end{equation}
where $i$ labels the unit cell and $\alpha$ denotes the internal degrees of freedom. In momentum space, this amounts to a uniform shift of the crystal momentum $\mathbf{k} \rightarrow \mathbf{k} + \mathbf{q}$, where $\mathbf{q} = (\Phi/L_x) \mathbf{e}_x$. Equivalently, every component of the pairing field acquires the same phase gradient:
\begin{equation}
\Delta_{i\alpha, j\beta} \rightarrow \Delta_{i\alpha, j\beta} e^{-2i \mathbf{q} \cdot \frac{\mathbf{r}_{i\alpha} + \mathbf{r}_{j\beta}}{2}}.
\end{equation}
To derive the free energy, we begin with the explicit form of the multiband Bogoliubov-de Gennes (BdG) Hamiltonian under the phase twist $\mathbf{q}$. In the Nambu spinor basis $\Psi_{\mathbf{k}} = (\psi_{\mathbf{k}}, u_T^\dagger \psi_{-\mathbf{k}}^*)^T$, where $u_T$ is the unitary part of the time-reversal operator, the BdG matrix is defined as:
\begin{equation}
\mathcal{H}_{\mathbf{k},\mathbf{q}} = \begin{pmatrix} h_{\mathbf{k}+\mathbf{q}} & \Delta_{\mathbf{k}} \\ \Delta_{\mathbf{k}}^\dagger & -u_T^\dagger h_{-\mathbf{k}+\mathbf{q}}^* u_T \end{pmatrix}.
\label{eq:BdG_matrix}
\end{equation}
Here, $h_{\mathbf{k}+\mathbf{q}}$ is the normal-state single-particle Hamiltonian block describing $M$ degrees of freedom including spin, and $\Delta_{\mathbf{k}}$ is the pairing matrix. Due to the fermion anticommutation relations required to write the Hamiltonian in this Nambu form, a trace of the normal state block emerges. The second-quantised mean-field Hamiltonian is given by:
\begin{equation}
H_{\text{mf}}(\mathbf{q}) = \frac{1}{2} \sum_{\mathbf{k}} \Psi_{\mathbf{k}}^\dagger \mathcal{H}_{\mathbf{k},\mathbf{q}} \Psi_{\mathbf{k}} + \frac{1}{2} \sum_{\mathbf{k}} \text{tr}(h_{\mathbf{k}+\mathbf{q}}) + E_{\text{const}},
\end{equation}
where $\text{tr}$ denotes the trace over the $M \times M$ single-particle degrees of freedom. $E_{\text{const}}$ contains the constant terms originating from the mean-field decoupling of the interactions (proportional to $\Delta_{\mathbf{k}}^\dagger \Delta_{\mathbf{k}}$) and the chemical potential.
The Helmholtz free energy per unit cell at inverse temperature $\beta = (k_B T)^{-1}$ is given by:
\begin{equation}
F(\mathbf{q}) = \frac{1}{2N_k} \sum_{\mathbf{k}} \text{Tr} \left[ \xi(\mathcal{H}_{\mathbf{k},\mathbf{q}}) \right] + \frac{1}{2N_k} \sum_{\mathbf{k}} \text{tr}(h_{\mathbf{k}+\mathbf{q}}) + E_{\text{const}}.
\label{eq:free_energy}
\end{equation}
Here, the matrix thermodynamic function is $\xi(\mathcal{H}) = -\beta^{-1} \ln(1 + \exp(-\beta \mathcal{H}))$, $\text{Tr}$ is the full matrix trace over the $2M \times 2M$ Nambu space, and $N_k$ is the number of momentum points.
The superfluid stiffness tensor is defined as the second derivative of the free energy with respect to the twist: $D_{s,\mu\nu} = \partial_{q_\mu}\partial_{q_\nu} F(\mathbf{q})|_{\mathbf{q}=0}$. 
First, we note that the scalar derivative of the thermodynamic function $\xi(E)$ yields the Fermi-Dirac distribution, $\xi'(E) = f(E) \equiv (1 + \exp(\beta E))^{-1}$. 
Working in the Hamiltonian eigenbasis $\mathcal{H}_{\mathbf{k},\mathbf{q}} |n\rangle = E_n |n\rangle$ and using the Feynman-Hellmann theorem, the first derivative is:
\begin{equation}
\partial_{q_\mu} F(\mathbf{q}) = \frac{1}{2N_k} \sum_{\mathbf{k}} \text{Tr} \left[ f(\mathcal{H}_{\mathbf{k},\mathbf{q}}) \left(\partial_{q_\mu} \mathcal{H}_{\mathbf{k},\mathbf{q}}\right) \right] + \frac{1}{2N_k} \sum_{\mathbf{k}} \text{tr}(\partial_{q_\mu} h_{\mathbf{k}+\mathbf{q}}).
\end{equation}
To evaluate the second derivative, we apply the product rule. Crucially, for a general multiband system, the Hamiltonian $\mathcal{H}_{\mathbf{k},\mathbf{q}}$ and its spatial derivative $\partial_{q_\mu} \mathcal{H}_{\mathbf{k},\mathbf{q}}$ do not commute. Therefore, we evaluate the formal variation of the Fermi matrix function $\partial_{q_\nu} f(\mathcal{H}_{\mathbf{k},\mathbf{q}})$ strictly within the eigenbasis using the divided difference of the eigenvalues:
\begin{equation}
\langle m | \partial_{q_\nu} f(\mathcal{H}_{\mathbf{k},\mathbf{q}}) | n \rangle = \frac{f(E_m) - f(E_n)}{E_m - E_n} \langle m | \partial_{q_\nu} \mathcal{H}_{\mathbf{k},\mathbf{q}} | n \rangle.
\end{equation}
Applying this to the trace and evaluating at $\mathbf{q}=0$ splits the stiffness into the diamagnetic response $\widetilde{D}_{\mu\nu}$ and the paramagnetic response $\chi^P_{\mu\nu}$, giving the standard linear response form $D_{s,\mu\nu} = \widetilde{D}_{\mu\nu} - \chi^P_{\mu\nu}$, where:
\begin{align}
\widetilde{D}_{\mu\nu} &= \frac{1}{2N_k} \sum_{\mathbf{k}} \sum_n f(E_n) \langle n | \partial^2_{q_\mu q_\nu} \mathcal{H}_{\mathbf{k},0} | n \rangle + \frac{1}{2N_k} \sum_{\mathbf{k}} \text{tr}(\partial^2_{q_\mu q_\nu} h_{\mathbf{k}}) \label{eq:diamagnetic_stiffness}\\
\chi^P_{\mu\nu} &= -\frac{1}{2N_k} \sum_{\mathbf{k}} \sum_{n,m} \frac{f(E_n) - f(E_m)}{E_n - E_m} \langle m | \partial_{q_\nu} \mathcal{H}_{\mathbf{k},0} | n \rangle \langle n | \partial_{q_\mu} \mathcal{H}_{\mathbf{k},0} | m \rangle \label{eq:paramagnetic_stiffness}
\end{align}
For intraband transitions and in case of degeneracies ($E_n \rightarrow E_m$), the divided difference ratio analytically converges to the derivative of the Fermi function $-\beta f(E_n)(1 - f(E_n))$.
Note that in \eqref{eq:diamagnetic_stiffness} and \eqref{eq:paramagnetic_stiffness}, the derivatives of the BdG Hamiltonian act exclusively on the normal-state blocks. We define the normal-state velocity matrix $v_\mu(\mathbf{k}) = \partial_{k_\mu} h_{\mathbf{k}}$ and inverse mass matrix $w_{\mu\nu}(\mathbf{k}) = \partial^2_{k_\mu k_\nu} h_{\mathbf{k}}$. For a generic time-reversal symmetry broken Hamiltonian with conserved $S_z$ and singlet pairing, the $2M \times 2M$ BdG matrix block-diagonalises into two decoupled $M \times M$ sectors. In the up-electron/down-hole sector spanned by the Nambu spinor $\Psi_{\mathbf{k}} = (c_{\mathbf{k}\uparrow}, c_{-\mathbf{k}\downarrow}^\dagger)^T$ with all orbitals besides spin invariant under time-reversal, the derivatives at $\mathbf{q}=0$ are:
\begin{align}
\partial_{q_\mu} \mathcal{H}_{\mathbf{k},0}^{\uparrow\downarrow} &= \begin{pmatrix} v_{\mu\uparrow}(\mathbf{k}) & 0 \\ 0 & -v_{\mu\downarrow}^*(-\mathbf{k}) \end{pmatrix} \\
\partial^2_{q_\mu q_\nu} \mathcal{H}_{\mathbf{k},0}^{\uparrow\downarrow} &= \begin{pmatrix} w_{\mu\nu\uparrow}(\mathbf{k}) & 0 \\ 0 & -w_{\mu\nu\downarrow}^*(-\mathbf{k}) \end{pmatrix}
\end{align}
The full Bogoliubov-de Gennes Hamiltonian possesses an artificial particle-hole redundancy, which doubles the degrees of freedom. By separating the system into decoupled sectors based on the conserved spin projection, we isolate these degrees of freedom. The matrices for the two sectors are related by charge conjugation symmetry, meaning the spectrum of the up-electron/down-hole block is mapped directly to the spectrum of the negative of the down-electron/up-hole block at $-\mathbf{k}$. Consequently, the isolated block $\mathcal{H}_{\mathbf{k},0}^{\uparrow\downarrow}$ does not exhibit internal particle-hole symmetry. While time-reversal invariant systems guarantee that the eigenvalues of a given block occur in $\pm E_n$ pairs, broken time-reversal symmetry means $E_n(\mathbf{k})$ and $-E_n(\mathbf{k})$ are not guaranteed to be in the same eigenspectrum. This absence of internal particle-hole symmetry mandates that the summation index $n$ in the trace equations must run over all $M$ eigenstates of the block, rather than summing only over the positive energy states as is done in the standard, time-reversal invariant BCS formulation. When summing the traces over both spin sectors, the $\frac{1}{2}$ Nambu-doubling prefactor in \eqref{eq:free_energy} is exactly canceled by the equal contribution from the down-electron/up-hole sector.
If the bare Hamiltonian preserves time-reversal symmetry, the relation $h_\downarrow(\mathbf{k}) = h_\uparrow^*(-\mathbf{k})$ further simplifies the operators to:
\begin{align}
\partial_{q_\mu} \mathcal{H}_{\mathbf{k},0}^{\uparrow\downarrow} &= \begin{pmatrix} v_{\mu\uparrow}(\mathbf{k}) & 0 \\ 0 & v_{\mu\uparrow}(\mathbf{k}) \end{pmatrix} = v_{\mu\uparrow}(\mathbf{k}) \otimes \sigma_0 \\
\partial^2_{q_\mu q_\nu} \mathcal{H}_{\mathbf{k},0}^{\uparrow\downarrow} &= \begin{pmatrix} w_{\mu\nu\uparrow}(\mathbf{k}) & 0 \\ 0 & -w_{\mu\nu\uparrow}(\mathbf{k}) \end{pmatrix} = w_{\mu\nu\uparrow}(\mathbf{k}) \otimes \sigma_z
\end{align}
For a pair-density wave (PDW) state ordering at a time-reversal invariant momentum (TRIM) $\mathbf{Q}$, the pairing term couples fermions at $\mathbf{k}$ and $-\mathbf{k}+\mathbf{Q}$. Because $2\mathbf{Q}$ is a reciprocal lattice vector, the Nambu basis can be constructed simply by shifting the hole operators without enlarging the unit cell: $\Psi_{\mathbf{k}}(\mathbf{Q}) = (c_{\mathbf{k}\uparrow}, c_{-\mathbf{k}+\mathbf{Q}\downarrow}^\dagger)^T$. The form of the BdG block matrix remains identically structured to the uniform $\mathbf{q}=0$ case, substituting only the shifted single-particle hole Hamiltonian $-h_{-\mathbf{k}+\mathbf{Q}+\mathbf{q}}^*$. Correspondingly, the implementation of the stiffness traces carries over directly by evaluating the normal-state hole velocity and inverse mass matrices at $-\mathbf{k}+\mathbf{Q}$, such that the derivatives of the block Hamiltonian are:
\begin{align}
\partial_{q_\mu} \mathcal{H}_{\mathbf{k},0}^{\uparrow\downarrow}(\mathbf{Q}) &= \begin{pmatrix} v_{\mu\uparrow}(\mathbf{k}) & 0 \\ 0 & -v_{\mu\downarrow}^*(-\mathbf{k}+\mathbf{Q}) \end{pmatrix} \\
\partial^2_{q_\mu q_\nu} \mathcal{H}_{\mathbf{k},0}^{\uparrow\downarrow}(\mathbf{Q}) &= \begin{pmatrix} w_{\mu\nu\uparrow}(\mathbf{k}) & 0 \\ 0 & -w_{\mu\nu\downarrow}^*(-\mathbf{k}+\mathbf{Q}) \end{pmatrix}
\end{align}

\section{Exact cancellation of diamagnetic and paramagnetic current response of flat band eigenstates}\label{app:darkloc}

The current response to a uniform vector potential 
\begin{equation}
j_{q=0,x}(\omega)=\left[\left\langle \widetilde{D}-\chi_{jj}(q=0,\omega)\right\rangle \right]A_{q=0}(\omega)
\end{equation}
 for an electron-pair-hopping Hamiltonian $H=\sum_{k\alpha\beta}b_{k\alpha}^{\dagger}\gamma_{k\alpha\beta}b_{k\beta}$
is defined by the diamagnetic $\widetilde{D}$ and paramagnetic $\chi_{jj}$
response functions
\begin{align}
j & =\frac{2e}{c}\sum_{k}b_{k\alpha}^{\dagger}\partial_{k}\gamma_{k\alpha\beta}b_{k\beta}\\
\widetilde{D} & =\left(\frac{2e}{c}\right)^{2}\sum_{k}b_{k\alpha}^{\dagger}\partial_{k}^{2}\gamma_{k\alpha\beta}b_{k\beta}\\
\chi_{jj} & =i\int_{t}e^{i\omega t}\left\langle \left[e^{-iHt}je^{iHt},j\right]\right\rangle 
\end{align}
Taking the expectation value in a particular eigenstate $|qm\rangle\equiv U_{qm\alpha}b_{q\alpha}^{\dagger}|0\rangle$
of $H$, where $U$ are the unitary matrices that diagonalise the
Hamiltonian matrix $U^{\dagger}\gamma U=\epsilon_{m}\delta_{mn}$, we find
\begin{align}
\left\langle \widetilde{D}\right\rangle  & =\left(\frac{2e}{c}\right)^{2}\sum_{k}\left\langle b_{k\alpha}^{\dagger}UU^{\dagger}\partial_{k}^{2}\gamma_{k\alpha\beta}UU^{\dagger}b_{k\beta}\right\rangle \\
 & =\left(\frac{2e}{c}\right)^{2}\sum_{k}\left\langle b_{k\alpha}^{\dagger}U_{k\alpha m}U_{km\alpha'}^{\dagger}\partial_{k}^{2}\gamma_{k\alpha'\beta'}U_{k\beta'm'}U_{km'\beta}^{\dagger}b_{k\beta}\right\rangle \\
 & =\left(\frac{2e}{c}\right)^{2}U_{qm\alpha'}^{\dagger}\partial_{k}^{2}\gamma_{k\alpha'\beta'}U_{q\beta'm}
\end{align}
and 
\begin{align}
\left\langle \chi_{jj}(\omega)\right\rangle  & =\left(\frac{2e}{c}\right)^{2}\sum_{m\neq n}\left[\frac{2}{\omega+i0^{+}-\left(\epsilon_{qm}-\epsilon_{qn}\right)}\right]\left|U_{qm\alpha'}^{\dagger}\partial_{q}\gamma_{q\alpha'\beta'}U_{q\beta'n}\right|^{2}.
\end{align}
This holds for a generic Hamiltonian. We now consider specifically
the matrix elements of the current operators for the pair-hopping
Hamiltonian on the checkerboard graph 
\begin{align}
\gamma & =\left(\begin{array}{cc}
2\cos k_{x} & 4\cos\frac{k_{x}}{2}\cos\frac{k_{y}}{2}\\
4\cos\frac{k_{x}}{2}\cos\frac{k_{y}}{2} & 2\cos k_{y}
\end{array}\right)\\
 & =\left(\cos k_{x}+\cos k_{y}\right)+\left(\cos k_{x}-\cos k_{y}\right)\tau_{z}+4\cos\frac{k_{x}}{2}\cos\frac{k_{y}}{2}\tau_{x}\\
\partial_{k_{x}}\gamma & =-\sin k_{x}\left(\tau_{z}+1\right)-2\sin\frac{k_{x}}{2}\cos\frac{k_{y}}{2}\tau_{x}\\
\partial_{k_{x}}^{2}\gamma & =-\cos k_{x}\left(\tau_{z}+1\right)-\cos\frac{k_{x}}{2}\cos\frac{k_{y}}{2}\tau_{x}
\end{align}
Let us consider the response of a flat band eigenstate at arbitrary
$q\neq M$: $|\psi_{q<}\rangle=U_{q<\alpha}b_{k\alpha}^{\dagger}|0\rangle$.
\begin{align}
\left\langle \widetilde{D}\right\rangle  & =\left(\frac{2e}{c}\right)^{2}U_{qm\alpha'}^{\dagger}\partial_{k}^{2}\gamma_{k\alpha'\beta'}U_{q\beta'm}\bigg|_{m=<}\\
 & =-\left(\frac{2e}{c}\right)^{2}U_{qm\alpha'}^{\dagger}\left(\cos k_{x}\left(\tau_{z}+1\right)+\cos\frac{k_{x}}{2}\cos\frac{k_{y}}{2}\tau_{x}\right)U_{q\beta'm}\Bigg|_{m=<}\\
\left\langle \chi_{jj}(\omega\to0)\right\rangle  & =\left(\frac{2e}{c}\right)^{2}\sum_{m\neq n}\left[\frac{-2}{\left(\epsilon_{qm}-\epsilon_{qn}\right)}\right]\left|U_{qm\alpha'}^{\dagger}\partial_{q}\gamma_{q\alpha'\beta'}U_{q\beta'n}\right|^{2}\bigg|_{m=<}\\
 & =\left(\frac{2e}{c}\right)^{2}\sum_{m\neq n}\left[\frac{-2}{\left(\epsilon_{qm}-\epsilon_{qn}\right)}\right]\left|U_{qm\alpha'}^{\dagger}\left(\sin k_{x}\left(\tau_{z}+1\right)+2\sin\frac{k_{x}}{2}\cos\frac{k_{y}}{2}\tau_{x}\right)U_{q\beta'n}\right|^{2}\bigg|_{m=<}
\end{align}

We find that the response is zero for all flat-band states except
at the M point where the band-touching renders the flat-band eigenvector
indeterminate.

\includegraphics[width=0.75\textwidth]{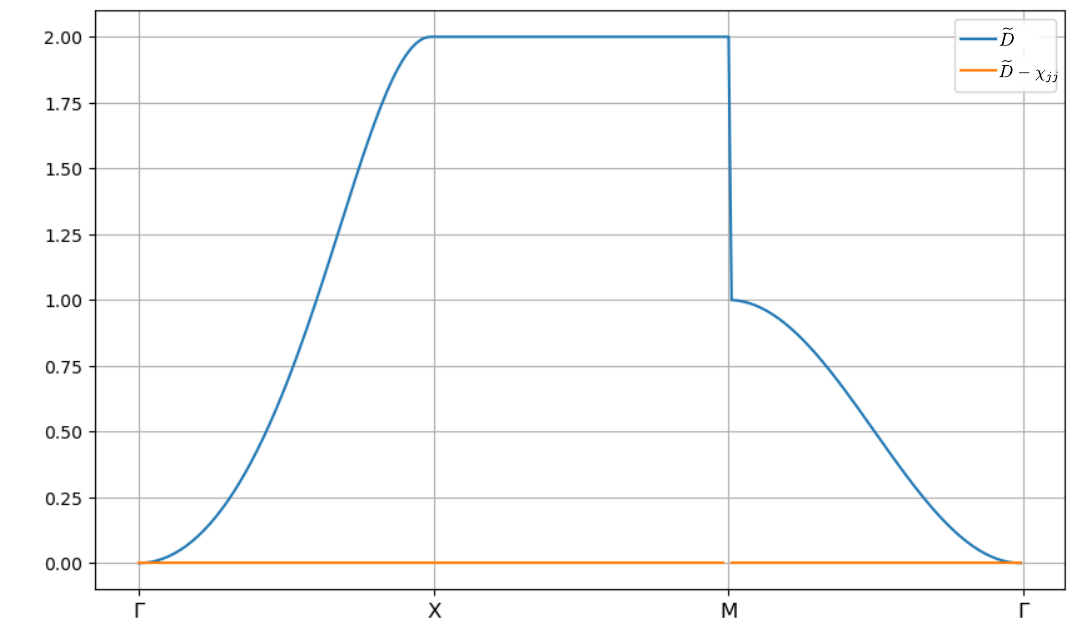}

This means that we can construct a localised eigenstate $|\Psi_{{\bf R},{\Phi}}\rangle\equiv\sum_{{\bf k}}e^{i{\bf k}\cdot{\bf R}}\phi_{{\bf k}}|\psi_{{\bf k}<}\rangle$,
which is also an eigenstate of the Hamiltonian. The current-response
of this state to a uniform vector potential is zero because the response
from individual crystal momenta decouple in the Kubo formula $\left\langle \psi_{r}\mid\widetilde{D}-\chi_{jj}(q=0,\omega)\mid\psi_{r}\right\rangle =\sum_{q}\langle q,-\mid\widetilde{D}-\chi_{jj}(q=0,\omega)\mid q,-\rangle=0$.

\section{Predictions for two-photoemission spectroscopy}\label{app:2eARPES}

The correlated emission of two photoelectrons in quantum materials
has a long and checkered history~\cite{berakdar1998,kouzakov2003,schumann2006,schumann2007,hattass2008,wehlitz2012,wallauer2012,jankala2014,huth2014,aliaev2018,trutzschler2017}. It is
often difficult to distinguish between the pre-existing correlations
between the low-energy electrons before interacting with the photon(s)
and the correlations imbued by the process of photoemission.
For instance, in the Auger process, a second photoelectron is emitted by the
radiation emitted by the filling of a core hole created by a photoemission~\cite{sawatzky1988,schumann2006,schumann2007,leitner2021}, 
so that the two emitted electrons are correlated even if their initial states were not. 
The other significant process by which one photon
results in the emission of two electrons is when an initial photoelectron
drags out another by a Coulomb drag or electron-energy loss process~\cite{mahmood2022}. Whether this one photon in-two electron out process can
in practice be disentangled from the two photon induced pair photoemission
process~\cite{devereaux2023} in practice is another open question~\cite{bonca2026}. 

Here, we point out that many of these murky issues are avoided in
the strong-pairing limit when the pair-binding energy is the largest
scale in the problem. Specifically, we consider the case of a fully
gapped superconductor whose pair-binding energy (2$\Delta$) minus
the thermal energy is greater than the photon energy minus the electron
work function. Since the photon cannot excite a single electron into
the delocalised states that extend to the detectors, there is only
one kind of initial state to consider, a two-particle bound state
that is excited into a two-electron extended state that reaches the
two detectors. Because the in-plane momentum carried by the photon
is negligible, the final center-of-mass momentum of the two photoelectrons
is equal to that of the initial two-electron bound state. In this
limit then, measuring the total energy and momentum of the two coincident
photoelectrons does amount to measuring the Cooper pair dispersion,
and thereby the superfluid stiffness directly from spectroscopy. \footnote{Note that this is not the regime in which experiments are currently
conducted; laser ARPES photon energy is typically 7eV, removing 5eV
as a typical work function of cuprate materials, still results in energies
one order of magnitude above the pairing gap ($2\Delta\approx100$ meV).}

For a material with fully gapped obstructed pairs, if the photon energy
is low enough, the multi-band dispersion with flat bands that is shown
in Fig. \ref{fig:checkerboard}(b) is observable in principle. Moreover,
the two-electron photoemission matrix element contains information
about the symmetry obstruction that defines the obstructed pairs.
\[
M_{2e}=\left\langle \psi_{{\rm final}}\left|{\bf A\cdot p}\right|\psi_{{\rm initial}}\right\rangle 
\]
To take the concrete example of the flat bands of the checkerboard,
the obstructed bosons are defined by their relative sign on the x-
and y-links, so that they are odd under the $\sigma_{d}$ mirror that
interchanges $x\leftrightarrow y$. If we consider photoelectrons
which are both emitted along $\Gamma M$, then the final state is
even under this symmetry, since both photoelectron wavefunctions must
not have nodes at the respective detectors. The initial state has
odd parity under $\sigma_{d}$ for all ${\bf q}\in\Gamma M$. Thus
the non-dispersive spectral weight in two-electron coincident photoemission
will be visible only if the photon polarisation is normal to the mirror
plane (ie. LH polarisation). Thus, the dipole selection rule of the
photoemission process can identify both the flat dispersion and the
symmetry character that defines obstructed pairs.

\section{Second-order perturbation theory for the low-energy effective Hamiltonians}\label{app:SW}

\subsection{Strong-coupling expansion for the attractive Hubbard model}\label{app:SW1}

We consider the attractive Hubbard model on the checkerboard graph
\begin{equation}
H = -\sum_{\langle i\alpha, j\beta\rangle,\sigma} \left( t_{\sigma} c_{i\alpha\sigma}^{\dagger}c_{j\beta\sigma}+h.c.\right) -U\sum_{i\alpha}
\left(c_{i\alpha\uparrow}^{\dagger}c_{i\alpha\uparrow}-\frac{1}{2}\right)
\left(c_{i\alpha\downarrow}^{\dagger}c_{i\alpha\downarrow}-\frac{1}{2}\right)
\end{equation}
where $i\alpha,j\beta$ are adjacent edges and $U>0$. In the strong-coupling limit ($|U| \gg t_\sigma$), we divide the Hilbert space into two sectors: a low-energy subspace (with projector $P_0$) of empty or doubly occupied sites, with energy per site $-|U|/4$, and an excited subspace (with projector $P_1$) of states with at least one singly occupied site, with energy per site $+|U|/4$. We will use degenerate perturbation theory to calculate the low energy effective Hamiltonian at second-order in $t/U$. 

To this end, we represent the fermions using standard Hubbard operators $X_{i\alpha}^{a\leftarrow b} = \ket{a}_{i\alpha}\bra{b}_{i\alpha}$. The physical states on each site are $\ket{0}, \ket{\uparrow}, \ket{\downarrow}, \text{and } \ket{2} = c_{\uparrow}^\dagger c_{\downarrow}^\dagger \ket{0}$. The fermion creation operators are mapped as:
\begin{align}
c_{i\alpha\uparrow}^\dagger &= X_{i\alpha}^{\uparrow\leftarrow 0} + X_{i\alpha}^{2\leftarrow\downarrow} \\
c_{i\alpha\downarrow}^\dagger &= X_{i\alpha}^{\downarrow\leftarrow 0} - X_{i\alpha}^{2\leftarrow\uparrow}
\end{align}
In terms of Hubbard operators, the unperturbed interaction Hamiltonian takes the form:
\begin{equation}
H_U = \frac{U}{4} \sum_{i\alpha} \left( X_{i\alpha}^{\uparrow\leftarrow\uparrow} + X_{i\alpha}^{\downarrow\leftarrow\downarrow} - X_{i\alpha}^{2\leftarrow 2} - X_{i\alpha}^{0\leftarrow 0} \right)
\end{equation}

The kinetic energy pertubation $H_K = -\sum_{\langle i\alpha, j\beta\rangle, \sigma} \left( t_\sigma c_{i\alpha\sigma}^\dagger c_{j\beta\sigma} + h.c. \right)$ can be decomposed into terms that change the total number of singly occupied sites by $+2, 0,$ and $-2$, denoted by $T_1, T_0,$ and $T_{-1}$ respectively:
\begin{equation}
H_K = T_1 + T_0 + T_{-1}
\end{equation}
Using the sign convention $\text{sgn}(\uparrow)=1$ and $\text{sgn}(\downarrow)=-1$, the operator that creates two single occupancies across a link is:
\begin{equation}
T_1 = -\sum_{\langle i\alpha, j\beta \rangle, \sigma} \text{sgn}(\sigma) \left( t_\sigma X_{i\alpha}^{\sigma\leftarrow 0} X_{j\beta}^{\bar{\sigma}\leftarrow 2} + t_\sigma^* X_{j\beta}^{\sigma\leftarrow 0} X_{i\alpha}^{\bar{\sigma}\leftarrow 2} \right)
\end{equation}
Its Hermitian conjugate, which destroys two single occupancies, is given by:
\begin{equation}
T_{-1} = \sum_{\langle i\alpha, j\beta \rangle, \sigma} \text{sgn}(\sigma) \left( t_\sigma^* X_{i\alpha}^{0\leftarrow\sigma} X_{j\beta}^{2\leftarrow\bar{\sigma}} + t_\sigma X_{j\beta}^{0\leftarrow\sigma} X_{i\alpha}^{2\leftarrow\bar{\sigma}} \right)
\end{equation}
The operator that preserves the number of single occupancies is:
\begin{equation}
T_0 = -\sum_{\langle i\alpha, j\beta \rangle, \sigma} \left( t_\sigma X_{i\alpha}^{\sigma\leftarrow 0} X_{j\beta}^{0\leftarrow\sigma} + t_\sigma X_{i\alpha}^{2\leftarrow\bar{\sigma}} X_{j\beta}^{\bar{\sigma}\leftarrow 2} + \text{h.c.} \right)
\end{equation}
By construction, these operators satisfy the commutation relations $[H_U, T_m] = m U T_m$.

\paragraph{The Schrieffer-Wolff transformation}
We construct a block-diagonal effective Hamiltonian $H_{\text{eff}} = P_0 e^{iS} H e^{-iS} P_0$ with a Hermitian generator $S = S^\dagger$, such that there are no off-diagonal matrix elements affected by the projector $P_0$ up to $\mathcal{O}(t^2/U)$. To eliminate the hopping to $\mathcal{O}(t/U)$, we require $[H_U, iS] + T_1 + T_{-1} = 0$. Using the commutators defined above, the required generator is:
\begin{equation}
iS = \frac{1}{U}(T_1 - T_{-1})
\end{equation}
Expanding the unitary transformation to second order, the effective Hamiltonian in the $P_0$ subspace becomes:
\begin{equation}
H_{\text{eff}} = P_0 \left( H_U + T_0 + \frac{1}{2}[iS, H_K] \right) P_0
\end{equation}
Since $H_U$ and $T_0$ both annihilate states in $P_0$, the surviving second-order term is:
\begin{equation}
H_{\text{eff}} = \frac{1}{2U} P_0 [T_1 - T_{-1}, T_1 + T_0 + T_{-1}] P_0 = -\frac{1}{U} P_0 T_{-1} T_1 P_0
\end{equation}

When expanding $T_{-1} T_1$, cross-terms between different links $\langle i\alpha, j\beta \rangle \neq \langle k\gamma, l\delta \rangle$ evaluate to zero in the $P_0$ subspace because they leave residual singly occupied sites. Thus, the effective Hamiltonian reduces to a sum over independent bonds:
\begin{equation}
H_{\text{eff}} = -\frac{1}{U} \sum_{\langle i\alpha, j\beta \rangle} P_0 T_{-1}^{(i\alpha, j\beta)} T_1^{(i\alpha, j\beta)} P_0
\end{equation}
We evaluate the product for a single link. The graded algebra of the Hubbard operators tracks the anti-commutation signs. Expanding the products and retaining only operators that transition within the $P_0$ subspace yields four non-vanishing components:
\begin{align*}
(X_{i\alpha}^{0\leftarrow\sigma'} X_{j\beta}^{2\leftarrow\bar{\sigma}'}) (X_{i\alpha}^{\sigma\leftarrow 0} X_{j\beta}^{\bar{\sigma}\leftarrow 2}) P_0 &= - \delta_{\sigma, \sigma'} X_{i\alpha}^{0\leftarrow 0} X_{j\beta}^{2\leftarrow 2} P_0 \\
(X_{j\beta}^{0\leftarrow\sigma'} X_{i\alpha}^{2\leftarrow\bar{\sigma}'}) (X_{j\beta}^{\sigma\leftarrow 0} X_{i\alpha}^{\bar{\sigma}\leftarrow 2}) P_0 &= - \delta_{\sigma, \sigma'} X_{i\alpha}^{2\leftarrow 2} X_{j\beta}^{0\leftarrow 0} P_0 \\
(X_{j\beta}^{0\leftarrow\sigma'} X_{i\alpha}^{2\leftarrow\bar{\sigma}'}) (X_{i\alpha}^{\sigma\leftarrow 0} X_{j\beta}^{\bar{\sigma}\leftarrow 2}) P_0 &= \delta_{\sigma', \bar{\sigma}} X_{i\alpha}^{2\leftarrow 0} X_{j\beta}^{0\leftarrow 2} P_0 \\
(X_{i\alpha}^{0\leftarrow\sigma'} X_{j\beta}^{2\leftarrow\bar{\sigma}'}) (X_{j\beta}^{\sigma\leftarrow 0} X_{i\alpha}^{\bar{\sigma}\leftarrow 2}) P_0 &= \delta_{\sigma', \bar{\sigma}} X_{i\alpha}^{0\leftarrow 2} X_{j\beta}^{2\leftarrow 0} P_0
\end{align*}
Applying the complex coefficients and summing over spin indices results in:
\begin{equation}
T_{-1}^{(i\alpha, j\beta)} T_1^{(i\alpha, j\beta)} P_0 = (|t_\uparrow|^2 + |t_\downarrow|^2) \left( X_{i\alpha}^{0\leftarrow 0} X_{j\beta}^{2\leftarrow 2} + X_{i\alpha}^{2\leftarrow 2} X_{j\beta}^{0\leftarrow 0} \right) + 2 t_\uparrow t_\downarrow X_{i\alpha}^{2\leftarrow 0} X_{j\beta}^{0\leftarrow 2} + 2 t_\uparrow^* t_\downarrow^* X_{i\alpha}^{0\leftarrow 2} X_{j\beta}^{2\leftarrow 0}
\end{equation}

We map the Hubbard operators acting on the $P_0$ subspace back to local electron pair field operators:
\begin{equation}
b_{i\alpha}^\dagger = c_{i\alpha\uparrow}^\dagger c_{i\alpha\downarrow}^\dagger = X_{i\alpha}^{2\leftarrow 0}, \quad b_{i\alpha} = c_{i\alpha\downarrow} c_{i\alpha\uparrow} = X_{i\alpha}^{0\leftarrow 2}
\end{equation}
The local pair number operator is $n_{i\alpha} = X_{i\alpha}^{2\leftarrow 2}$, while the empty state projector is $1 - n_{i\alpha} = X_{i\alpha}^{0\leftarrow 0}$. Utilising the shifted pair-density operator $\tilde{n}_{i\alpha} = n_{i\alpha} - \frac{1}{2}$, the diagonal part of the interaction becomes:
\begin{equation}
X_{i\alpha}^{0\leftarrow 0} X_{j\beta}^{2\leftarrow 2} + X_{i\alpha}^{2\leftarrow 2} X_{j\beta}^{0\leftarrow 0} = n_{i\alpha} + n_{j\beta} - 2n_{i\alpha}n_{j\beta} = \frac{1}{2} - 2\tilde{n}_{i\alpha}\tilde{n}_{j\beta}
\end{equation}
Substituting these identities back into $H_{\text{eff}}$ yields the hard-core boson Hamiltonian:
\begin{equation}
H_{\text{eff}} = \sum_{\langle i\alpha, j\beta \rangle} \left[ - \frac{2 t_\uparrow t_\downarrow}{U} b_{i\alpha}^\dagger b_{j\beta} - \frac{2 t_\uparrow^* t_\downarrow^*}{U} b_{j\beta}^\dagger b_{i\alpha} + \frac{2(|t_\uparrow|^2 + |t_\downarrow|^2)}{U} \tilde{n}_{i\alpha} \tilde{n}_{j\beta} \right]
\end{equation}
We identify the effective hopping amplitude $t_b = \frac{-2 t_\uparrow t_\downarrow}{U}$ and the repulsive pair interaction $V_b = \frac{2(|t_\uparrow|^2 + |t_\downarrow|^2)}{U}$.

\subsection{From hard-core bosons model to low-energy effective dimer model}\label{app:SW2}
We consider the resulting hard-core boson model restricted to a single square plaquette of the lattice, consisting of four sites/rungs labeled $A$ (top), $B$ (right), $C$ (bottom), and $D$ (left). The system is analysed in the strong-coupling limit $V_b \gg |t_b|$. The unperturbed Hamiltonian is the nearest-neighbour repulsion $H_V = V_b (n_A n_B + n_B n_C + n_C n_D + n_D n_A)$, and the perturbation is the complex hopping $H_K = t_b (b_A^\dagger b_B + b_B^\dagger b_C + b_C^\dagger b_D + b_D^\dagger b_A) + \text{h.c.}$ 

The total number of bosons is conserved, and we restrict the system to 2 bosons. The low-energy subspace $P_0$ consists of the two classical dimer coverings with zero adjacent bosons ($E_0 = 0$):
\begin{equation}
\ket{\hdimer} = b_A^\dagger b_C^\dagger \ket{0}, \quad \ket{\vdimer} = b_B^\dagger b_D^\dagger \ket{0}
\end{equation}
The excited subspace $P_1$ consists of four states, each containing exactly one pair of adjacent bosons ($E_1 = V_b$). These represent dimer violations at the four vertices:
\begin{equation}
\ket{\violI} = b_C^\dagger b_D^\dagger \ket{0}, \quad \ket{\violII} = b_A^\dagger b_D^\dagger \ket{0}, \quad \ket{\violIII} = b_B^\dagger b_C^\dagger \ket{0}, \quad \ket{\violIV} = b_A^\dagger b_B^\dagger \ket{0}
\end{equation}

Applying the Schrieffer-Wolff transformation at second order in $t_b$ gives the effective Hamiltonian in the $P_0$ subspace:
\begin{equation}
H_{\text{eff}} = P_0 V_{\text{hop}} \frac{1}{E_0 - H_0} V_{\text{hop}} P_0 = -\frac{1}{V_b} P_0 V_{\text{hop}}^2 P_0
\end{equation}

We evaluate the action of $V_{\text{hop}}$ on the horizontal dimer state $\ket{\hdimer}$:
\begin{align}
V_{\text{hop}} \ket{\hdimer} &= \left( t_b^* b_B^\dagger b_A + t_b b_D^\dagger b_A + t_b b_B^\dagger b_C + t_b^* b_D^\dagger b_C \right) b_A^\dagger b_C^\dagger \ket{0} \\
&= t_b^* b_B^\dagger b_C^\dagger \ket{0} + t_b b_D^\dagger b_C^\dagger \ket{0} + t_b b_A^\dagger b_B^\dagger \ket{0} + t_b^* b_A^\dagger b_D^\dagger \ket{0} \\
&= t_b^* \ket{\violIII} + t_b \ket{\violI} + t_b \ket{\violIV} + t_b^* \ket{\violII}
\end{align}
Applying $V_{\text{hop}}$ a second time and projecting onto $\ket{\vdimer}$, we sum the transition amplitudes from each of the four intermediate states. For example, the transition from $\ket{\violI}$ requires creating $B$ and destroying $C$, yielding $V_{\text{hop}} (t_b \ket{\violI}) = (t_b b_B^\dagger b_C) (t_b b_C^\dagger b_D^\dagger \ket{0}) = t_b^2 \ket{\vdimer}$. Evaluating all paths yields the total off-diagonal matrix element:
\begin{equation}
\bra{\vdimer} H_{\text{eff}} \ket{\hdimer} = -\frac{1}{V_b} \left[ (t_b^*)^2 + t_b^2 + t_b^2 + (t_b^*)^2 \right] = -\frac{2(t_b^2 + (t_b^*)^2)}{V_b}
\end{equation}
Similarly, projecting back onto $\ket{\hdimer}$ yields the diagonal contribution. For instance, returning from $\ket{\violIII}$ requires $b_A^\dagger b_B$, giving $(t_b b_A^\dagger b_B) (t_b^* b_B^\dagger b_C^\dagger \ket{0}) = |t_b|^2 \ket{\hdimer}$. Summing all four paths gives:
\begin{equation}
\bra{\hdimer} H_{\text{eff}} \ket{\hdimer} = -\frac{1}{V_b} \left( |t_b|^2 + |t_b|^2 + |t_b|^2 + |t_b|^2 \right) = -\frac{4|t_b|^2}{V_b}
\end{equation}
By symmetry, the remaining diagonal element is $\bra{\vdimer} H_{\text{eff}} \ket{\vdimer} = -4|t_b|^2/V_b$. 

The second-order perturbation lowers the energy of all states in $P_0$. The final effective model is:
\begin{equation}
H = -\frac{2(t_b^2 + (t_b^*)^2)}{V_b} \left(\ket{\hdimer}\bra{\vdimer}+ \text{h.c.}\right) - \frac{4|t_b|^2}{V_b} \left(\ket{\vdimer}\bra{\vdimer} + \ket{\hdimer}\bra{\hdimer} \right)
\end{equation}
where setting $t_b = i |t_b|$ yields the plaquette Hamiltonian 
\begin{equation}
H = -\frac{4|t_b|^2}{V_b} \left(\ket{\hdimer}\bra{\vdimer}+ \bra{\hdimer}\bra{\vdimer}\right) - \frac{4|t_b|^2}{V_b} \left(\ket{\vdimer}\bra{\vdimer} + \ket{\hdimer}\bra{\hdimer} \right)\label{eq:RK1app}
\end{equation}
which is the negative of the usual dimer Hamiltonian at the Rokhsar-Kivelson point.

\section{Corrections to mean-field superfluid stiffness from gap phase relaxation}\label{app:iDelta}

We do not find any \textcolor{black}{purported} corrections to the mean-field superfluid stiffness of the form
\begin{equation}
D_s^{(corr)} - D_s^{(0)} = - \mathbf{b}^T H^{-1} \mathbf{b}
\end{equation}
discussed by Huhtinen \textit{et al.} \cite{Huhtinen2022}, where $D_s^{(0)} = \partial^2 \Omega / \partial A^2$ is second derivative of the grand canonical potential $\Omega$ with respect to a twist in the boundary condition implemented by a vector potential $A$, $b_\alpha = \partial^2 \Omega / \partial A \partial \Delta_\alpha^I$ is the mixed derivative vector of $\Omega$ with respect to $A$ and the imaginary part of the gap $\Delta^I$, and $H$ is the Hessian of $\Omega$ with respect to the gap phases ($H_{\alpha\beta} = \partial^2 \Omega / \partial \Delta_\alpha^I \partial \Delta_\beta^I$).

We understand the absence of such corrections as a consequence of inversion symmetry in our model. Specifically, under a spatial inversion operation $\mathcal{P}: \mathbf{r} \to -\mathbf{r}$ centered around a checkerboard lattice site, the uniform vector potential transforms as $\mathbf{A} \to -\mathbf{A}$, while the mean-field ground state and order parameters remain invariant ($\Delta_\alpha^I \to \Delta_\alpha^I$). Since the grand canonical potential $\Omega(\mathbf{A}, \mathbf{\Delta}^I)$ is invariant under spatial inversion, it requires $\Omega(\mathbf{A}, \mathbf{\Delta}^I) = \Omega(-\mathbf{A}, \mathbf{\Delta}^I)$. The mixed derivative $b_\alpha$ corresponds to the coefficient of the cross-term $A \Delta_\alpha^I$ in the Taylor expansion of $\Omega$. As $A$ is odd and $\Delta_\alpha^I$ is even under inversion, this odd cross-term must vanish ($b_\alpha = -b_\alpha = 0$). Consequently, this gap phase relaxation correction is identically zero, \textcolor{black}{and the mean-field superfluid stiffness is given as usual by $D_s^{(0)} = \partial^2 \Omega / \partial A^2$.}

Beyond mean-field, the superfluid stiffness is given by the response to a static \emph{non-uniform} vector potential in the transverse limit ${\bf q}_\perp \to 0$ which physically corresponds to the linear response to a inversion-even magnetic field, rather than an inversion-odd electric field. Within mean-field theory, these two response functions are both equal to the response to a static uniform vector potential ${\bf A}(q=0,\omega=0)$. As is well known, the finite value of this response function is an artefact of the mean-field approximation and appropriately accounting for Gaussian corrections to the saddle-point action restores this $q=0,\omega=0$ response to zero as required by gauge invariance~\cite{anderson1958b,ambegaokar1961a,paramekantiEffectiveActionsPhase2000}. 

\section{Robustness of obstructed pairs away from the time-reversal antisymmetric limit $t_{\uparrow}=-t_\downarrow$}\label{app:robustness}

\begin{figure}[h]
    \centering
    \includegraphics[width=0.6\linewidth]{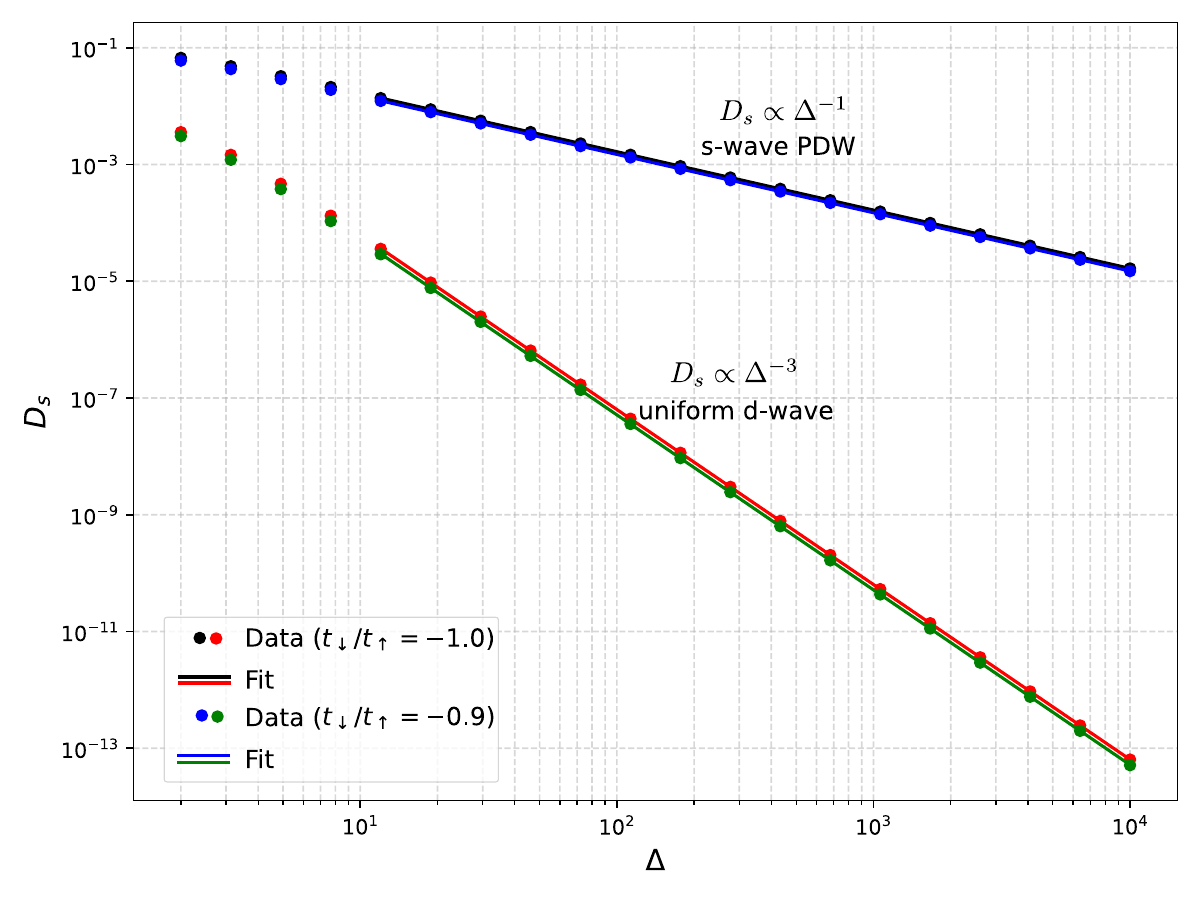}
    \caption{\textbf{Stiffness scaling away from the antisymmetric limit:} Strong-pairing asymptote of the superfluid stiffness at $t_\downarrow/t_\uparrow = -0.9$ and $-1.0$, showing the robust $\Delta^{-3}$ scaling for uniform d-wave (obstructed) pairs.}
    \label{fig:scaling_tdn_m0p9}
\end{figure}

\begin{figure}[h]
    \centering
    \includegraphics[width=0.9\linewidth]{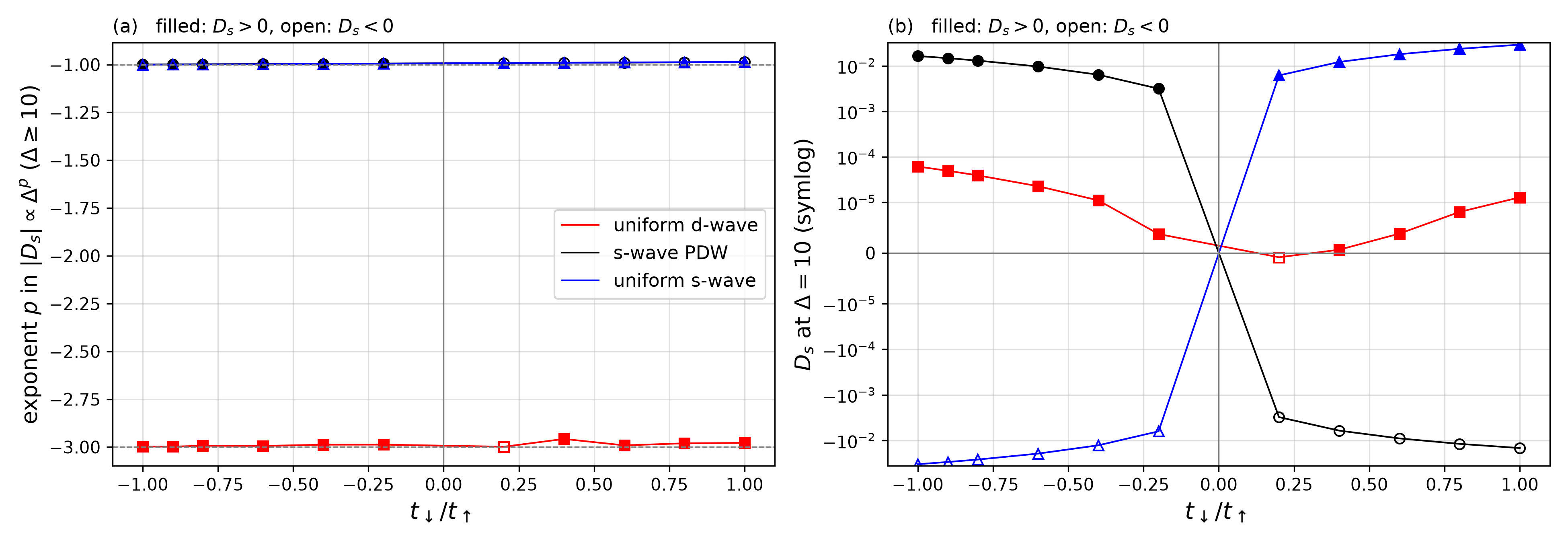}
    \caption{\textbf{Superfluid stiffness across the parameter space:} (a) The asymptotic scaling exponent $p$ (where $D_s \propto \Delta^p$) remains $p\approx-3$ for obstructed pairs across the full parameter range. (b) Magnitude and sign of $D_s$ at $\Delta=10t_\uparrow$. Open symbols correspond to negative stiffness, indicating that the corresponding state is unstable (e.g., uniform s-wave pairing when $t_\downarrow/t_\uparrow < 0$).}
    \label{fig:exponent_phase_space}
\end{figure}

To demonstrate that the anomalous $D_s \sim \Delta^{-3}$ scaling for obstructed pairs is not fine-tuned to the exactly time-reversal antisymmetric point $t_\uparrow = -t_\downarrow$, we evaluate the superfluid stiffness away from this limit. Fig.~\ref{fig:scaling_tdn_m0p9} confirms that the $\Delta^{-3}$ scaling holds identically at $t_\downarrow/t_\uparrow = -0.9$. Fig.~\ref{fig:exponent_phase_space} provides a finer sampling across $t_\downarrow/t_\uparrow \in [-1, 1]$, demonstrating that the obstructed pairs have positive stiffness throughout the entire regime where the effective pair-hopping is positive ($t_\uparrow t_\downarrow < 0$).

\end{appendix}





\bibliography{dwaveRVB}


\end{document}